\documentclass[lettersize,journal]{IEEEtran}
\usepackage{amsmath,amssymb,amsthm}
\usepackage[linesnumbered,ruled]{algorithm2e}
\usepackage{array}
\usepackage[caption=false,font=normalsize,labelfont=sf,textfont=sf]{subfig}
\usepackage{textcomp}
\usepackage{stfloats}
\usepackage{url}
\usepackage{verbatim}
\usepackage{subfig}
\usepackage{graphicx}
\usepackage{multirow} 
\usepackage{caption}
\usepackage{cite}
\usepackage{diagbox}
\newtheorem{theorem}{Theorem}[section]

\title{Optimization of BLE Broadcast Mode in Offline Finding Network}

\author{L Zhang, C Feng\, T Xia\IEEEauthorrefmark{2}}

\begin{document}
\maketitle

\begin{abstract}
In the Offline Finding Network(OFN), offline Bluetooth tags broadcast to the surrounding area, the finder devices receiving the broadcast signal and upload location information to the IoT(Internet of Things) cloud servers, thereby achieving offline finding of lost items. This process is essentially a Bluetooth low energy (BLE) neighbor discovery process(NDP). In the process, the variety of Bluetooth scan modes caused by the scan interval and scan window settings affects the discovery latency of finder devices finding the tag broadcast packets. To optimize the experience of searching for lost devices, we propose the CPBIS-mechanism, a certain proportion broadcast-intervals screening mechanism that calculates the most suitable two broadcast intervals and their proportion for offline tags. This reduces discovery latency in the BLE NDP, improves the discovery success rate, further enhances the user experience. To our knowledge, we are the first to propose a comprehensive solution for configuring the broadcast interval parameters of advertisers in BLE NDP, particularly for configurations involving two or more broadcast intervals. We evaluated the results obtained by CPBIS on the nRF52832 chip. The data shows that the CPBIS-mechanism achieves relatively low discovery latencies for multiple scan modes.
\end{abstract}


\section{Introduction}

Various attempts have been made to facilitate the retrieval of lost items in daily life\cite{zekavat2021overview}. A typical way is to binding easily lost items with offline BLE tags to determine their locations. Products of this type include Samsung’s Smart Tag, Apple’s Air Tag, and Huawei’s HUAWEI Tag\cite{smarttag,airtagwebsite,HUAWEITag}, all of which operate mainly based on Bluetooth Low Energy\cite{tosi2017performance}. They work roughly as follows, lost offline tag sends BLE advertisement packets to the surroundings to enable finder devices(working as BLE scan devices) to discover it. The presence of offline tags and numerous finder devices, with the assistance of an IoT(Internet of Things) cloud servers, constitutes an Offline Finding Network (OFN)\cite{li2023design}. The framework of OFN is shown as Fig.1.

\begin{figure}
    \centering
    \includegraphics[width=0.8\linewidth]{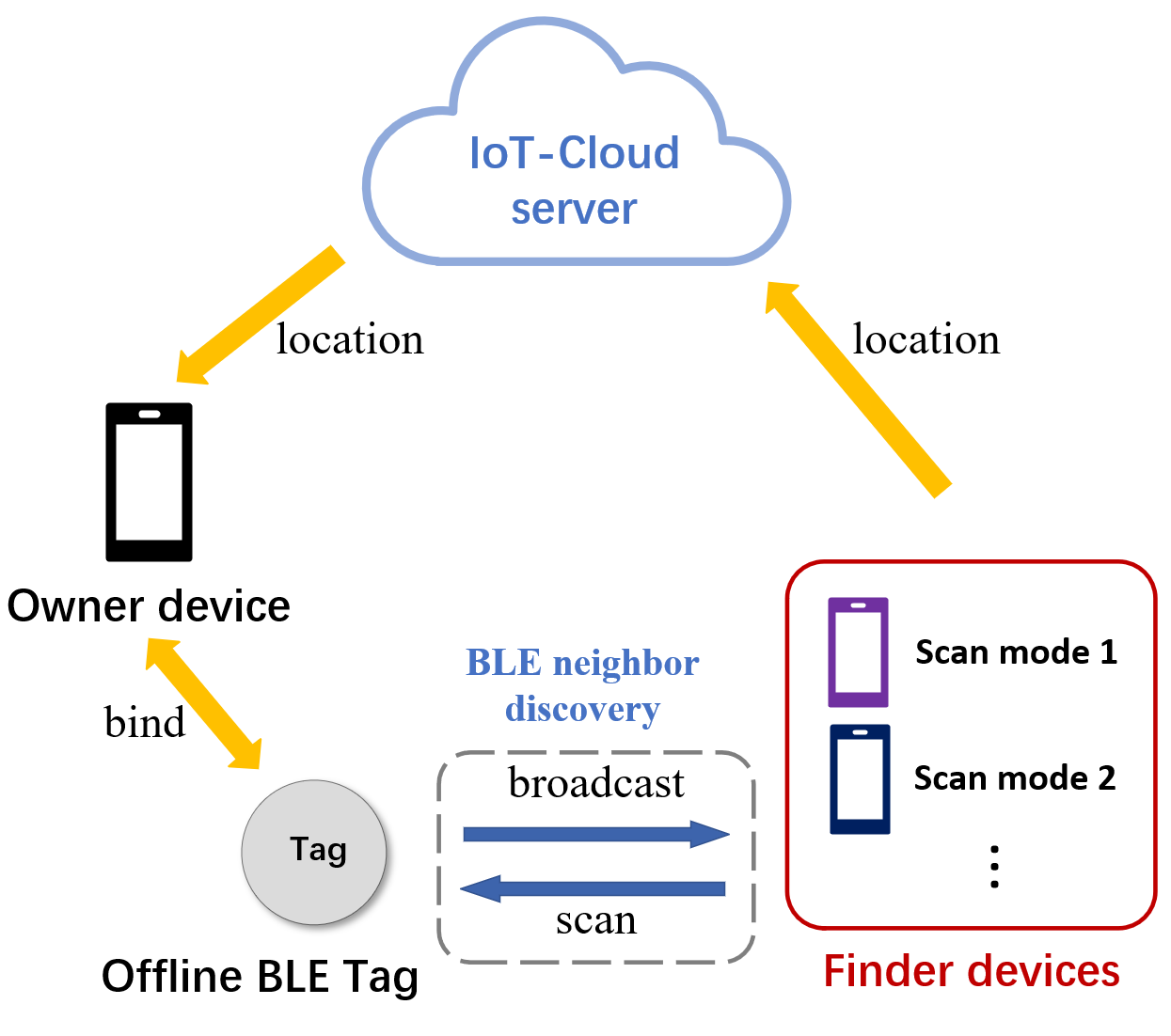}
    \caption{Offline Finding Network}
    \label{fig:enter-label}
\end{figure}
An important indicator that impacts the success rate of offline finding is the time it takes for finder devices around the tag to discover it. Referred to the discovery latency of the BLE NDP, which is the time from a finder device enters the broadcast range of the offline tag until it fully receives an advertisement packet sent by the tag. A lower discovery latency implies a higher discovery success rate.

Several studies have found that discovery latency can be reduced by adjusting some parameters of the process\cite{luo2019ble,shan2018advertisement}. Variables involved in BLE NDP include the broadcast mode of offline tag, the scan modes of the finder devices, and the time a pedestrian carrying a finder device passes through the broadcast range of the lost offline tag. Among these influencing factors, the broadcast mode of BLE tag and the scan mode of the finder device are controllable parameters. However, owing to significant variations in power and brand among finder devices, there are numerous types of scan modes and adjustments to them will be complicated. Therefore, changing the broadcast mode of the BLE tag is a better option for decreasing the discovery latency of BLE NDP. In this paper, we choose two broadcast intervals broadcast mode based on consideration of tag power consumption and discovery latency. 

Our paper propose a certain proportion broadcast intervals screening mechanism applied in OFN for multiple scan modes, which we called CPBIS-mechanism. This two broadcast interval screening mechanism allows finder devices in different scan modes to achieve relatively low discovery latencies.
\begin{figure}[h]
    \centering
    \includegraphics[width=0.4\linewidth]{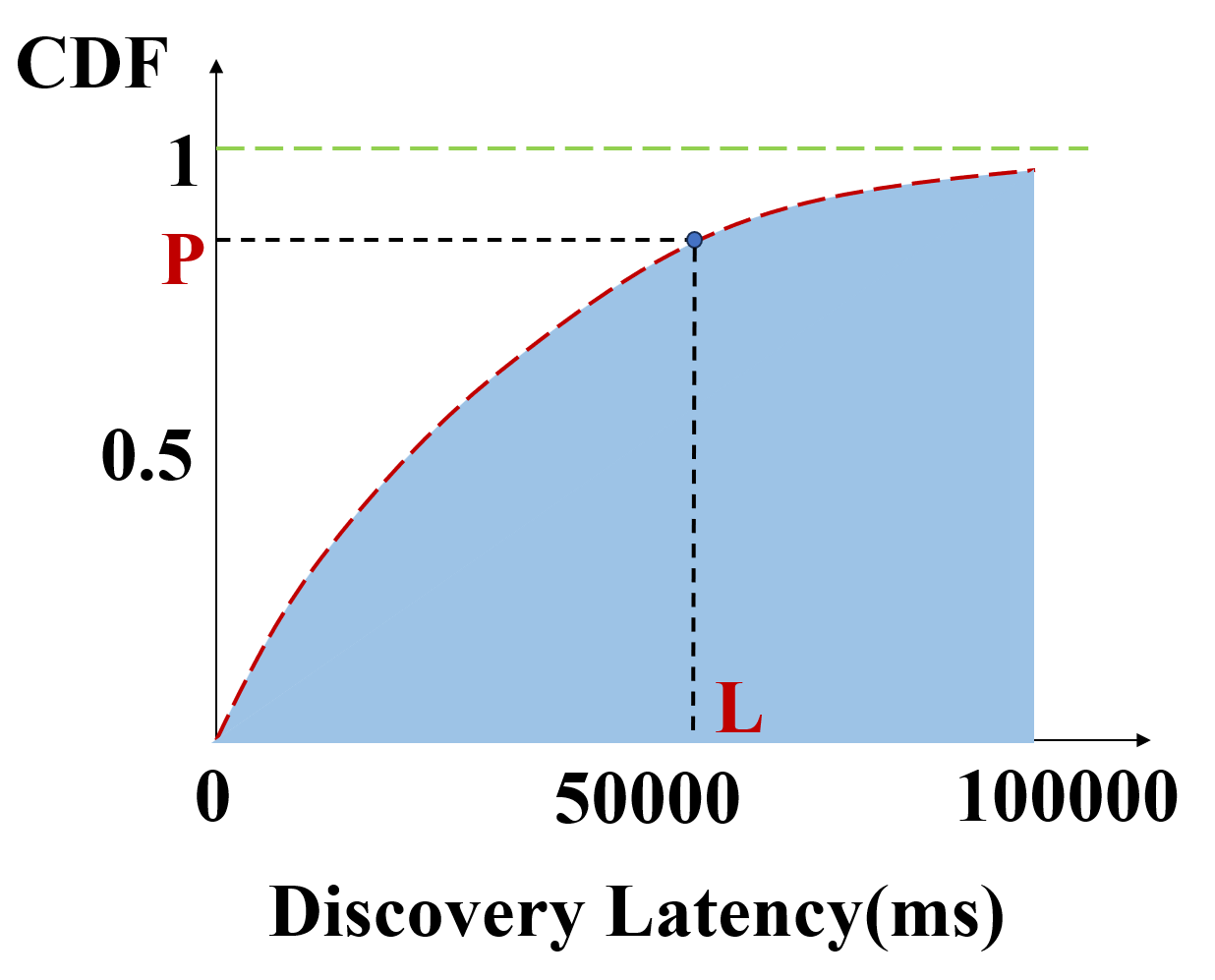}
    \caption{a CDF graph of discovery latency}
    \label{fig:enter-label}
\end{figure}

To optimize the discovery latency of BLE NDP by focusing on broadcast mode, it is necessary to understand how the setting of the broadcast interval affects the discovery latency in different scan modes. Using the BLE neighbor discovery simulator (Blender\cite{ding2022blender}), determine input parameters like scan interval(T), scan window(W), and broadcast interval(A), a graph of the cumulative distribution function (CDF) of the discovery latency can be derived shown as Fig.2. Selecting a probability $P$ will corresponds to a theoretical worst discovery latency $L$. By continuously increase the broadcast interval from a initial value,A series of corresponding worst discovery latency values will be obtained. Use the "broadcast interval-discovery latency" data pair sequence, we obtained the "broadcast interval-discovery latency" distributions of different scan modes. This is a waveform graph of the distribution of worst-case discovery latencies corresponding to the different broadcast interval for each scan mode. Superimpose the distributions of different scan modes and divide the waveform into peak-to-peak intervals, every trough point is a local optimum point. All trough points form a initial screened broadcast interval sequence. After the interval sequence is re-optimized, an ascending sequence of broadcast intervals will be obtained. With the power consumption of offline tag limited by manufacturer, which is reflected in the minimum equivalent broadcast interval $A_{min}$ of the tag (broadcast interval cannot be smaller than $A_{min}$ of the maximum power constraint.). Using $A_{min}$ as a dividing line, After a series of operations, CPBIS will eventually select out the optimal two broadcast intervals and their proportion, which is the final result.

The CPBIS-mechanism is a comprehensive "two-broadcast interval screening mechanism." It is capable of figuring out a two broadcast mode of the offline tag that is relatively optimized for all required scan modes. The contribution of the paper is threefold.

We designed a two-broadcast interval screening mechanism "CPBIS" for offline BLE tag in BLE NDP. At present, there is very little research in this area, and there is a lack of complete solutions for the parameter settings of BLE NDP. This screening mechanism can optimize the broadcast mode of offline tags in Offline finding network works for various scan modes existing in the market, achieving low discovery latency while reducing tag power consumption.

The CPBIS-mechanism analyzes different scan modes, obtaining the broadcast interval-discovery latency distribution. It superimposes the discovery latency distributions of different scan modes and finally selects the two most appropriate broadcast intervals and their proportion under the required $A_{min}$.

We evaluated the CPBIS-mechanism on the Nordic nRF52832 chip\cite{nRF52832chip}. While the scan modes are 5120ms/512ms and 4096ms/1024ms, each accounting for 50\%. The time limit for each BLE neighbor discovery is 40s. The weighted average discovery latency of the broadcast mode obtained by the CPBIS-mechanism is 10.595s. The weighted average discovery success rate is 98.3\%, which is the highest among the three compared broadcast modes.

\section{motivation}
\subsection{The possibility of controlling broadcast interval to reduce the discovery latency in BLE NDP}

A significant factor affects the success rate of finding the lost offline Tag is the discovery latency of BLE NDP. According to previous research\cite{li2023design,kindt2019optimal}, in the BLE NDP. For a certain scan mode(setting a scan mode by defining the scan interval and scan window), the "broadcast interval-discovery latency" distribution derived from CDF graphs of discovery latency is a waveform with peaks and troughs, shown as Fig.3. The relationship between the minimum worst discovery latency and the broadcast interval is a straight line with a positive slope. Therefore, we can select broadcast intervals that are close to the valley and as small as possible to reduce the discovery latency. Based on this, under the premise of the $A_{min}$, If a broadcast mode can be set to minimize the discovery latency. This will increase the discovery success rate and improve the user experience.
\begin{figure}[h]
    \centering
    \includegraphics[width=1.0\linewidth]{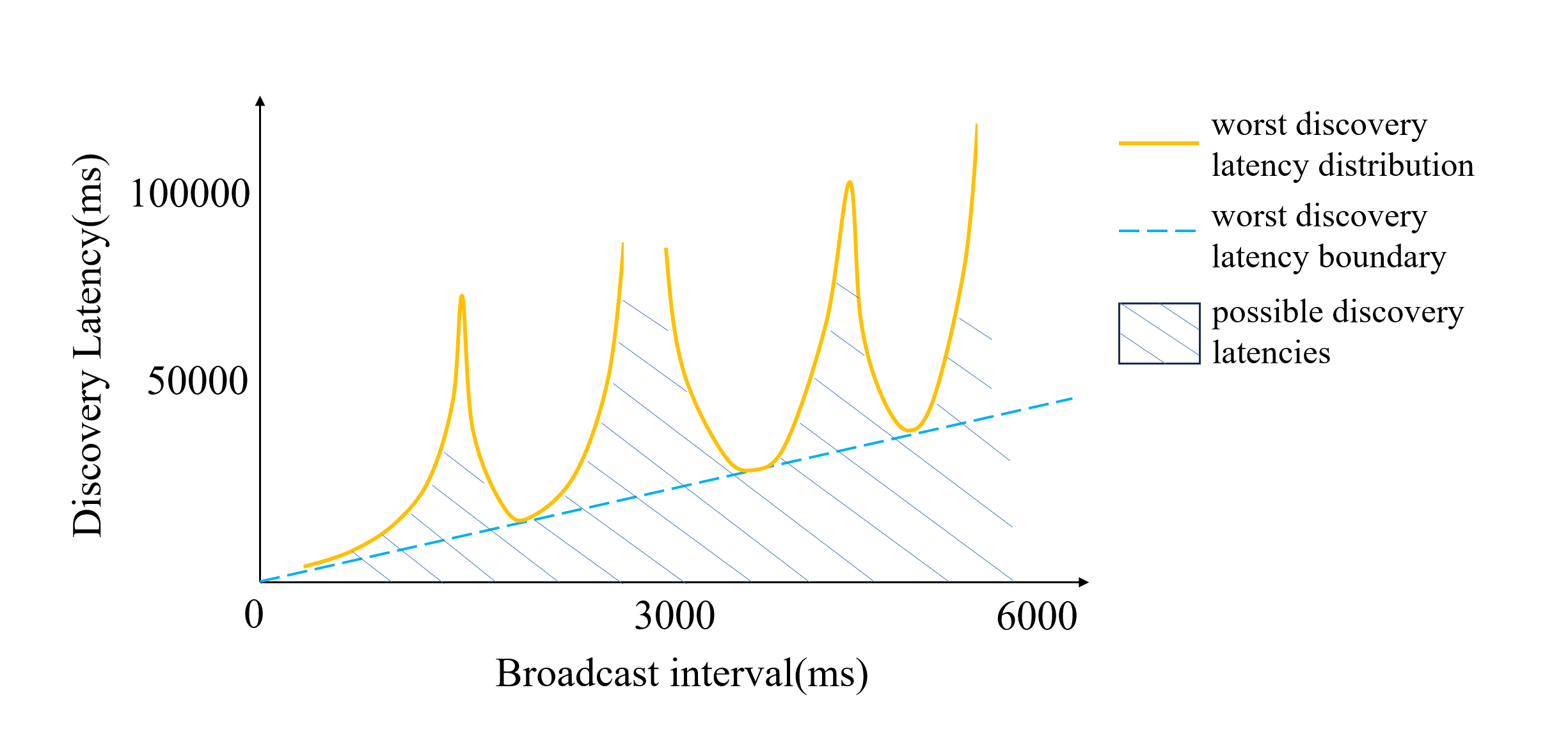}
    \caption{"broadcast interval-discovery latency" distribution}
    \label{fig:enter-label}
\end{figure}

\subsection{The necessity of setting two broadcast intervals}
To achieve a relatively low discovery latency with the existence of the $A_{min}$. Assume a short broadcast interval $A_1$ smaller than the $A_{min}$ and a long interval $A_2$ larger than $A_{min}$ in Fig.4. If only $A_1$ was selected, although the discovery latency can be lower, the power constraint is not satisfied. Besides, if only a long interval is chosen, the constraint is satisfied, but the discovery latency will be higher. Therefore, in contrast to selecting only one broadcast interval, the CPBIS-mechanism selects two intervals on each side of $A_{min}$ and control their proportions to make the equivalent broadcast interval meets the power constraint, CPBIS takes into account both the low discovery latency of the short broadcast interval and the low power consumption of the long broadcast interval. It can achieve a lower discovery latency than a single broadcast interval while satisfying the manufacturer's required power consumption constraints.
\begin{figure}[h]
    \centering
    \includegraphics[width=0.6\linewidth]{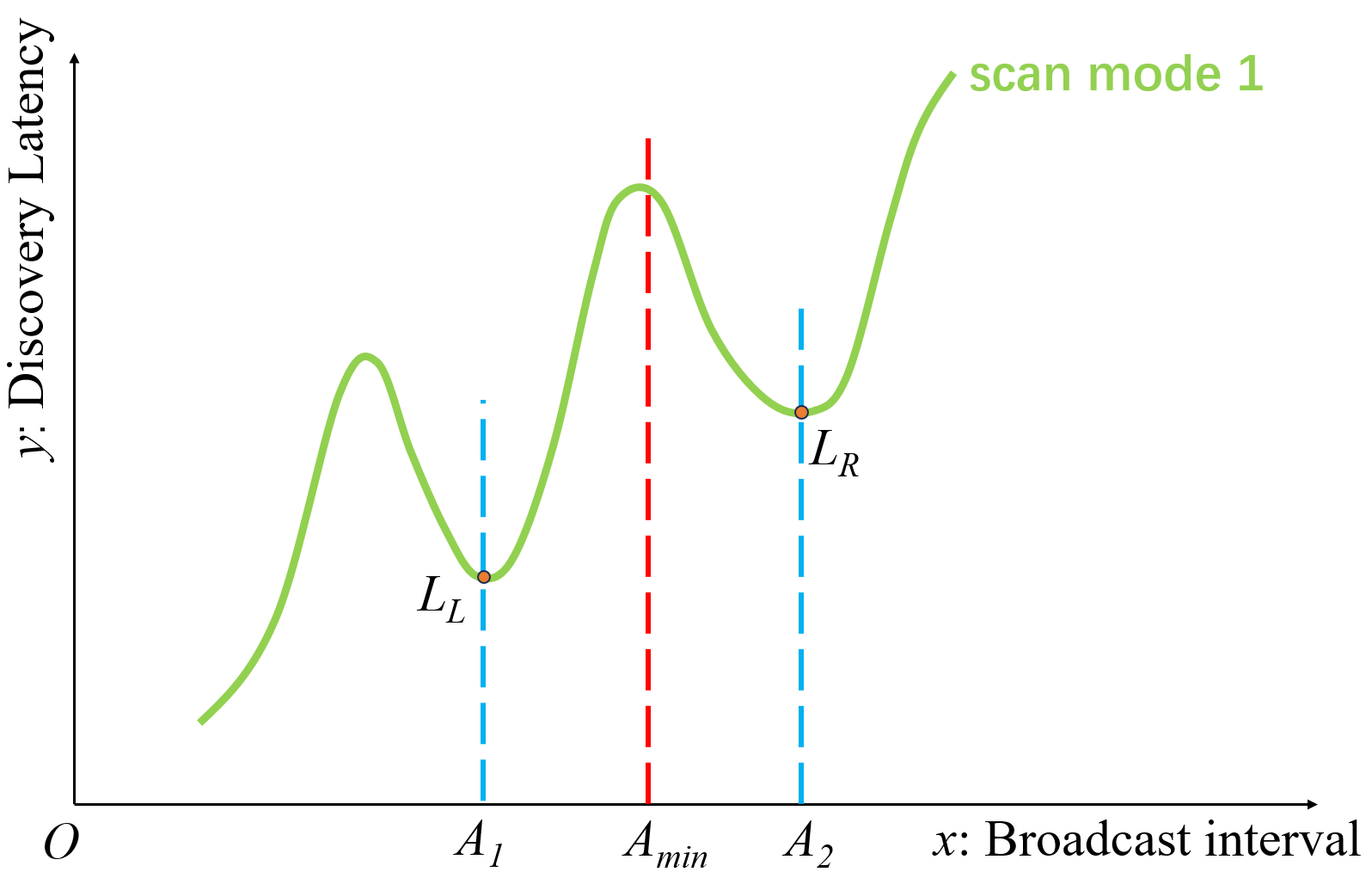}
    \caption{Two local optimum broadcast intervals within a scan mode,and $A_{min}$}
    \label{fig:enter-label}
\end{figure}
\section{Overview of CPBIS-mechanism and background}
\subsection{Offline Finding Network}
The background of the paper is OFN, OFN mainly consists of three parts: offline tags, IoT cloud servers, and finder devices. Take the Air Tag as an example, the tag does not carry a GPS chip or a networking module, which means positioning can be achieved offline.
\begin{figure}[h]
    \centering
    \includegraphics[width=0.8\linewidth]{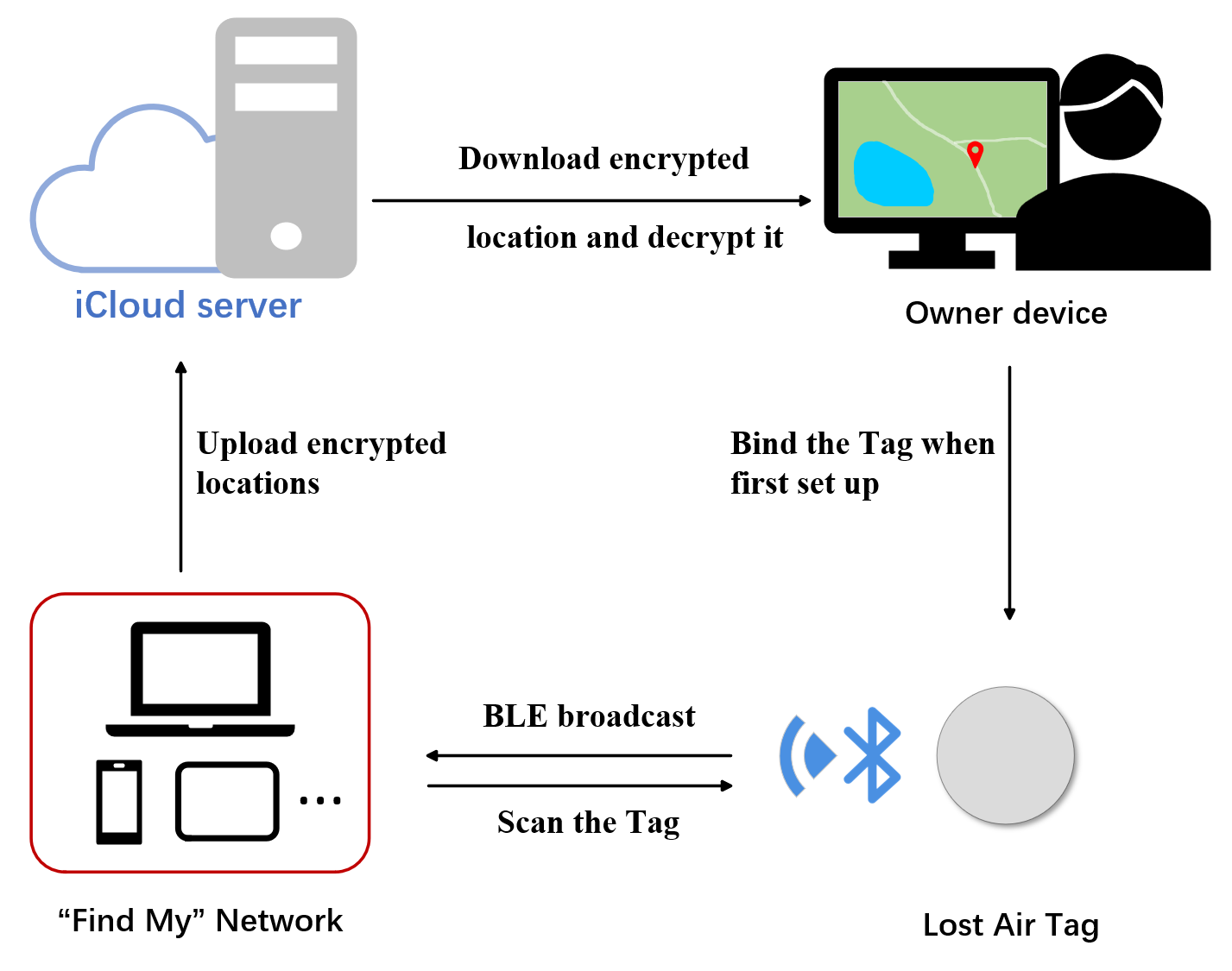}
    \caption{Offline Tag working mode}
    \label{fig:enter-label}
\end{figure}

The tag is a Bluetooth tracking device, mainly uses three technologies: BLE, UWB, and NFC. BLE is used to implement the offline finding function. The Tag can be attached to daily necessities such as bicycles and handbags to track the location of these items. The "Find My" network\cite{findmynet} helps locate the tag by detecting the Bluetooth signal of the lost Tag and transmitting the location back to its owner via iCloud\cite{findmysecurity}. The "Find My" network consists of finder devices like iPhones, iPads, and Macs from around the world. The specific working mode of the tag is: When the Tag is lost, it emits BLE advertisement to the surroundings. The signal will be detected by nearby finder devices, and an approximate location of the lost Air Tag will be sent to iCloud. The original owner of the item can see its location on the map in the "Find My" app. The entire process is anonymous and encrypted\cite{heinrich2021can}, as shown in Fig.5.

This article focuses on the BLE tag's offline finding feature. We are committed to reducing the BLE NDP discovery latency and improving the success rate of this process.

\subsection{Key parameters in CPBIS-mechanism}
In order to get the final two broadcast intervals and their proportions, some preparatory works needs to be done before. The first step is to obtain the "broadcast interval-discovery latency" distribution of a fixed scan mode. Using the simulator "blender", we obtained CDF distribution of the discovery latency under a certain scan mode ($T, W$) and a initial broadcast interval $A_1$, as shown in left of Fig.6. Find the discovery latency corresponding to the probability P, the latency $L_1$ with the interval $A_1$ form a data pair($A_1, L_1$), by continuously increase the value of the broadcast interval  in the simulator, a series of (A, L) data pair will be obtained. Use the "broadcast interval-discovery latency" data pair sequence, the discovery latency distribution will be obtained, like the right graph in Fig.6.
\begin{figure}[h]
    \centering
    \includegraphics[width=1.0\linewidth]{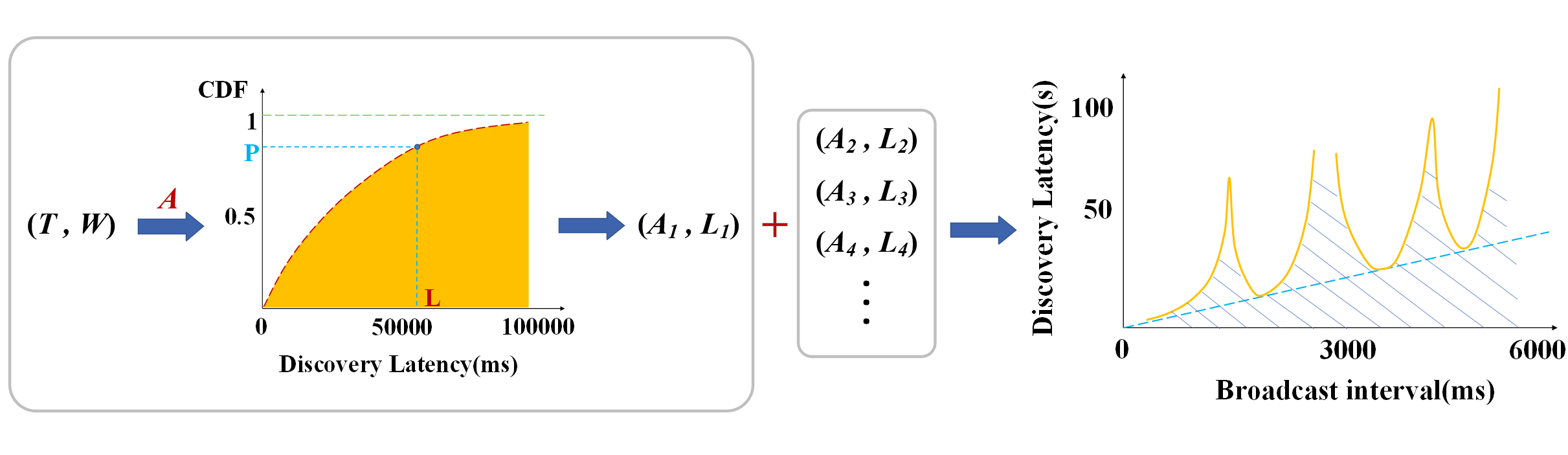}
    \caption{Use the (A, L) data pair sequence output by the simulator "blender" to construct a "broadcast interval-discovery latency" distribution}
    \label{fig:enter-label}
\end{figure}

Because of the existence of the $A_{min}$, the selection of the two broadcast intervals needs to satisfy the power constraint. First, the equivalent broadcast interval $\overline A$ calculated by the two broadcast intervals $A_1$ and $A_2$ is: $ \overline A = \delta A_1 + (1 -\delta) A_2 $, where the value of $\delta$ is proved by CPBIS and ranges from 0 to 1. The minimum equivalent broadcast interval constraint is: 
\begin{equation}
    \overline A \geq A_{min}
\end{equation}

Which means that the equivalent broadcast intervals of the two selected broadcast intervals cannot be smaller than $A_{min}$ of the maximum power constraint.
\begin{figure}[h]
    \centering
    \includegraphics[width=0.6\linewidth]{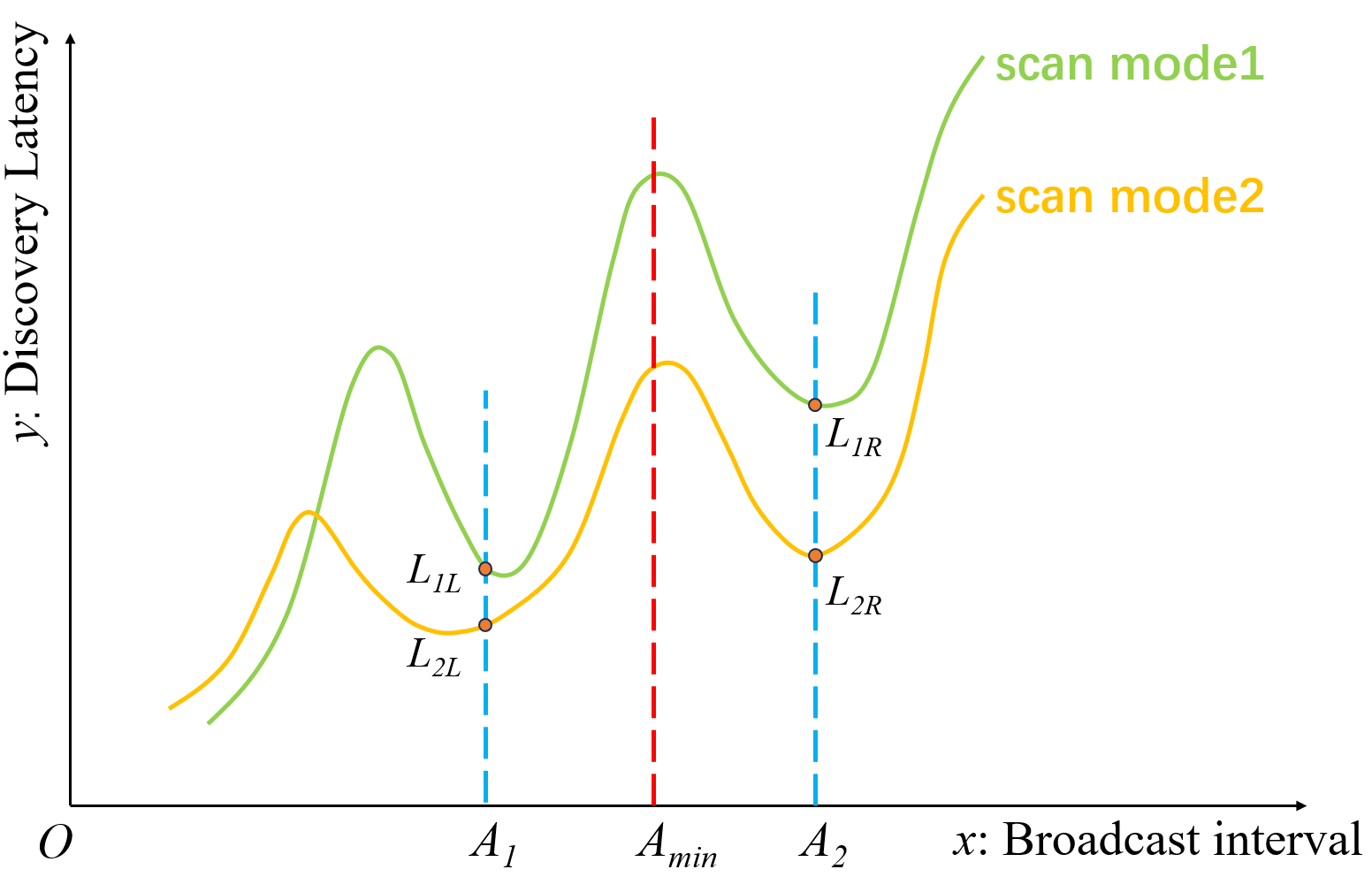}
    \caption{Discovery latency distribution of two scan modes and two selected broadcast intervals:$A_1,A_2$}
    \label{fig:enter-label}
\end{figure}

As shown in Fig.7, in the two-broadcast interval broadcast mode, $A_1$ and $A_2$ are the selected broadcast intervals. The green curve represents the discovery latency distribution of scan mode1, while the orange curve represents scan mode2. The discovery latency of scan mode 1 and scan mode 2 corresponding to A1 are$L_{1L}$ and $L_{2L}$, respectively, corresponding to A2 are$L_{1R}$ and $L_{2R}$, respectively. The market shares of the two scan modes are $\omega_1, \omega_2$ respectively. The two broadcast interval $A_1$ and $A_2$ respectively have a share of the broadcasting cycle c, 1 - c. The weighted average discovery latency is calculated as:
\begin{equation}
L = c(\omega_1 L_{1L} + \omega_2 L_{2L} ) + (1-c)(\omega_1 L_{1R} + \omega_2 L_{2R} )
\end{equation}

\subsection{CPBIS-mechanism framework}
This section introduces the role of CPBIS in OFN and outlines the main steps of CPBIS.

The CPBIS-mechanism in offline finding framework is shown as Fig.8. CPBIS recommends broadcast interval configurations based on various required scan modes. First, it obtains the “broadcast interval-discovery latency” distributions for different scan modes, and superimposes them to filter out the initial screened broadcast interval sequence consisting of all trough nodes. Then, the initial screened interval sequence will be optimized for several rounds, resulting in the final determination of the required two broadcast intervals and their proportions.
\begin{figure}[h]
    \centering
    \includegraphics[width=1.0\linewidth]{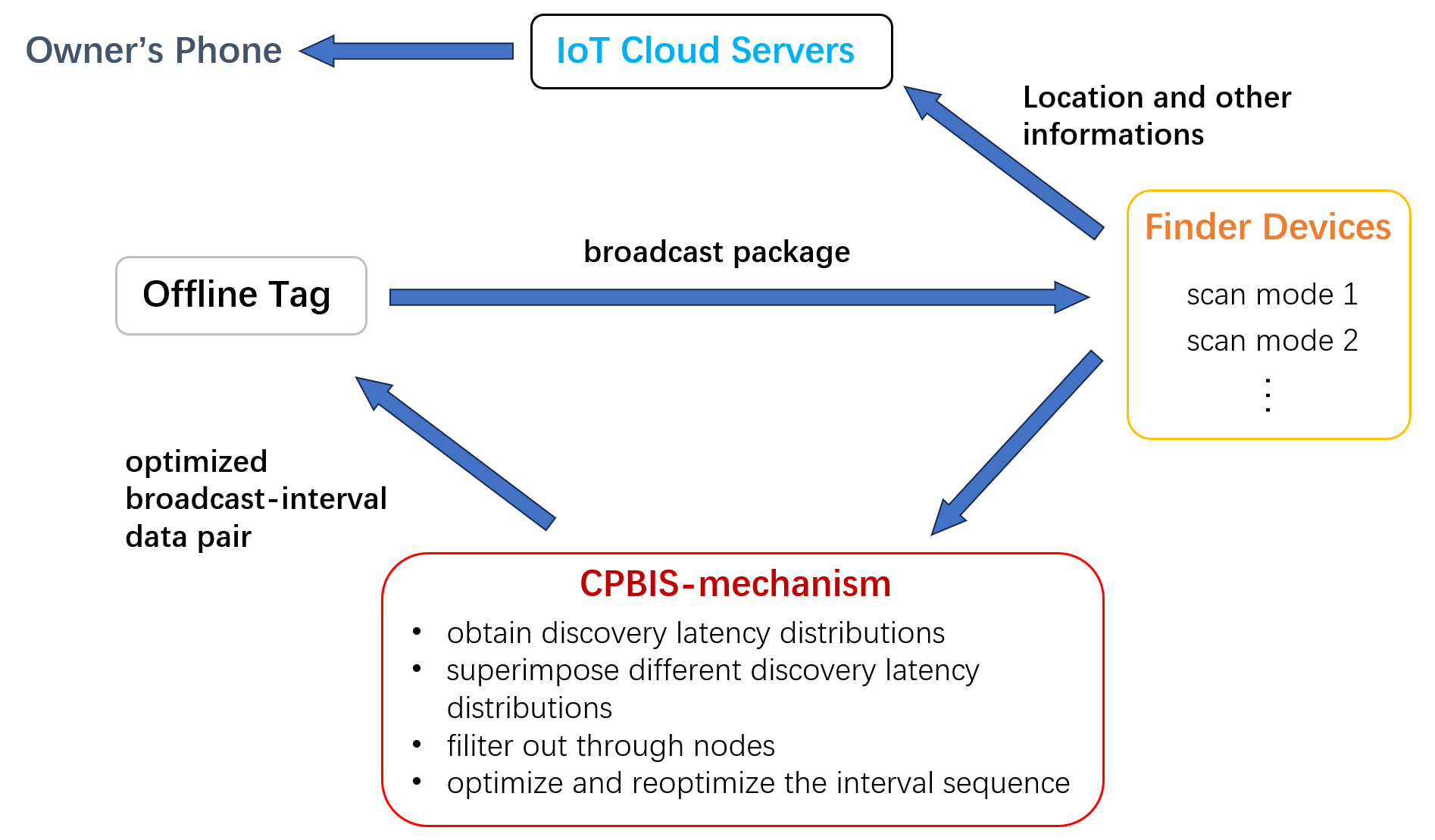}
    \caption{"CPBIS" in OFN}
    \label{fig:enter-label}
\end{figure}

\subsubsection{Obtain the broadcast interval-discovery latency distribution}
set the main input parameters, scan modes, a sequence of ascending broadcast intervals containing a left boundary and a right boundary. For a fixed scan mode $(T, W)$, as described in section B, constantly run the blender framework with increasing broadcast intervals in the ascending interval sequence. Use the ($A, L$) data pairs to form a discovery latency distribution belonging to the scan mode. Then, change the scan mode and repeat this part of the work to obtain the discovery latency distributions of different scan modes. As shown in the left figure of Fig.9(a), there are two "broadcast interval-discovery latency" distributions of two scan modes, indicated by the blue and orange lines, respectively.

\subsubsection{Initial screened broadcast interval sequence}
In order to ensure that the selected broadcast intervals enable each scan mode to have a relatively low discovery latency, it is necessary to simultaneously consider the extent to which the discovery latency of each scan mode is affected by the broadcast interval. we superimpose the discovery latency distributions of different scan modes by market share to form a hybrid discovery latency distribution graph. shown as the right figure of Fig.9(a). divide the superimposed graph into peak to peak intervals. The trough point between each two peaks is the optimal point in the interval, as the weighted average discovery latency generated by any point in the interval is higher than this point. The broadcast interval sequence composed of such local optimum points is the initial screened broadcast interval sequence. These points constitute the scatter plot on the left of Fig.9(b).
\begin{figure}[h]
    \centering
    \begin{minipage}{0.5\textwidth}
    \includegraphics[width=\linewidth]{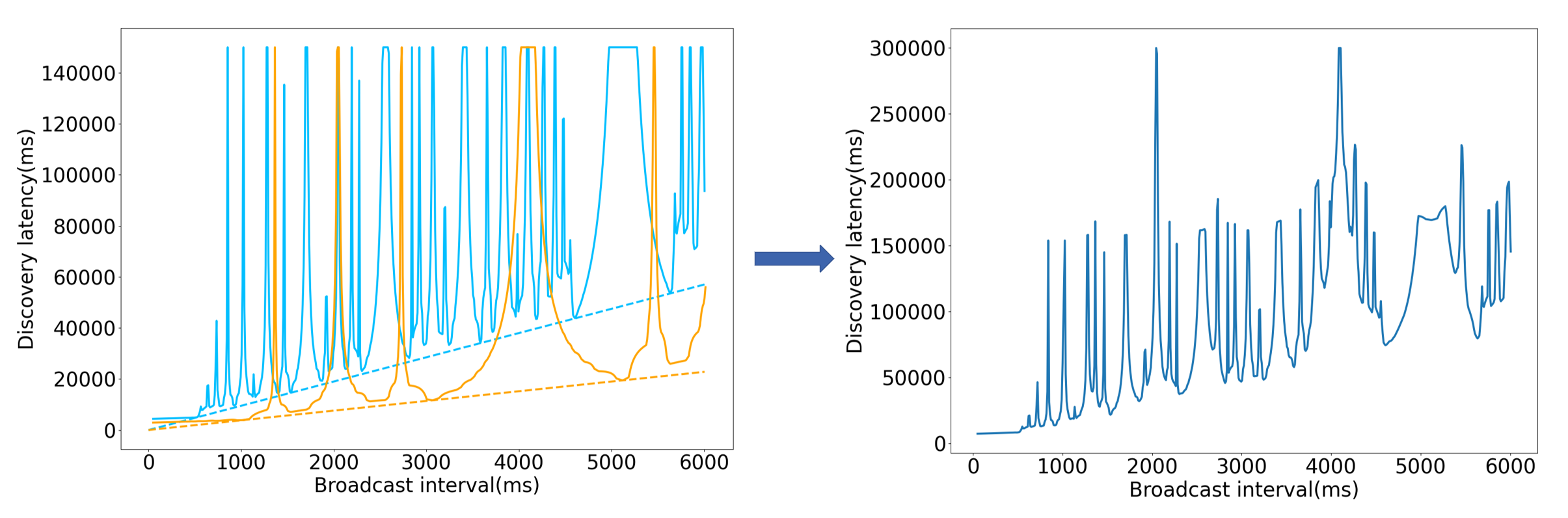}
    \caption*{(a):superimpose the discovery latency distributions}
    \label{}
    \end{minipage}\hfill
    \begin{minipage}{0.5\textwidth}
    \includegraphics[width=\linewidth]{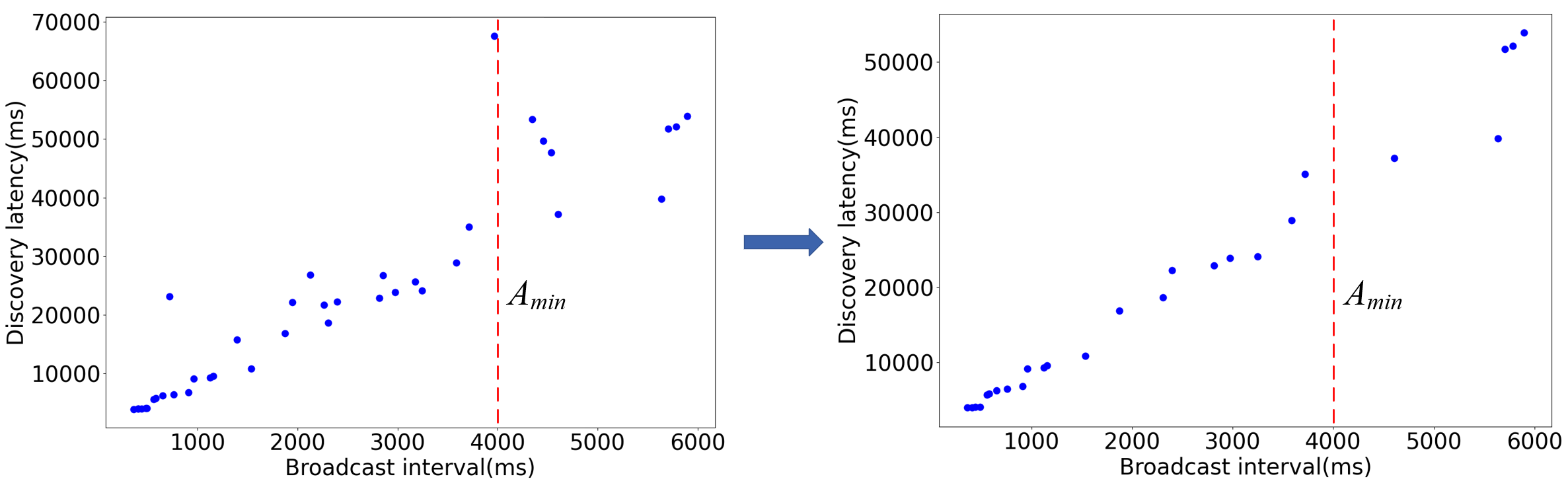}
    \caption*{(b):select the trough points and delete none-increasing points from the right side to left}
    \label{fig:enter-label}
    \end{minipage}\hfill
    \caption{the broadcast-interval screening process in the CPBIS-mechanism}
    \label{fig:multi-image}
\end{figure}
\subsubsection{Interval sequence re-optimization}
First, we define the non-increasing points. Elements whose discovery latency is larger than the first point to their right are termed non-increasing elements. eliminate non-increasing points from right to left to further screen the initial broadcast interval sequence obtained in the previous step. Since the selection of two broadcast intervals must satisfy the constraint in Equation(1). And the distribution of the minimum worst discovery latency in the discovery latency distribution graph satisfies a straight line with a positive slope of P·T/W, under the premise of selecting a broadcast interval on each side, there are two reasons for eliminating the non-increasing elements from the initially screened interval sequence.

First analyze the distribution of the two broadcast intervals to be selected. Since the trough points in the discovery latency distribution basically form an increasing straight line, the lower boundary is a straight line passing through the origin with a slope of P·T/W. For the selection of two broadcast intervals, taking $A_{min}$ as the dividing line,  if the two broadcast intervals are both on the left side of $A_{min}$, no matter how the proportion of the two intervals is adjusted, the constraint in equation (1) cannot be satisfied; if the two broadcast intervals are both on the right of the dividing line, although the constraint $\overline A \geq A_{min}$ is satisfied, fix either of the two broadcast intervals, an alternative broadcast interval of the another with same proportion satisfied the constraint can be found on the left side and the weighted average discovery latency is lower than this case; therefore, the two final broadcast intervals must be located on each side of $A_{min}$. 

Subsequently, analyze the reason for eliminating non-increasing elements in the initial screened sequence, whether on the left or right side of $A_{min}$, the non-increasing element has a smaller broadcast interval than the latter element, which means it consumes more power, but the discovery latency value is higher; in addition, for another broadcast interval adapted to this broadcast interval, the subsequent element of it must also be adapted (the constraint of equation(1) holds after replacing the non-increasing element by the element after it), and it works better, i.e., the resulting weighted average discovery latency is lower.

Eventually we obtained the re-optimized broadcast interval sequence in the scatter plot on the right of Fig.9(b).

\subsubsection{Local optimum interval data pair sequence}
For the re-optimized broadcast interval sequence obtained from the previous step, divide it into two parts by the $A_{min}$, start from the first point on the right side of $A_{min}$: select out the combination with the minimum discovery latency of all the points on the left side with this point. Until the last point on the right side is traversed, a local optimum broadcast interval data pair sequence is finally formed. The reason will be explained in the Detailed Design section.

\subsubsection{Optimal interval data pair}
In this step, we further research the local optimal interval data pair sequence obtained in the previous step, and filter out the broadcast interval data pair with the smallest weighted average discovery latency, which is the final result. The detailed selection of these two broadcast intervals will be explained in the Detailed Design section.

Fig.10 illustrates the five main working steps of CPBIS-mechanism and the output of each step.
\begin{figure}[h]
    \centering
    \includegraphics[width=0.9\linewidth]{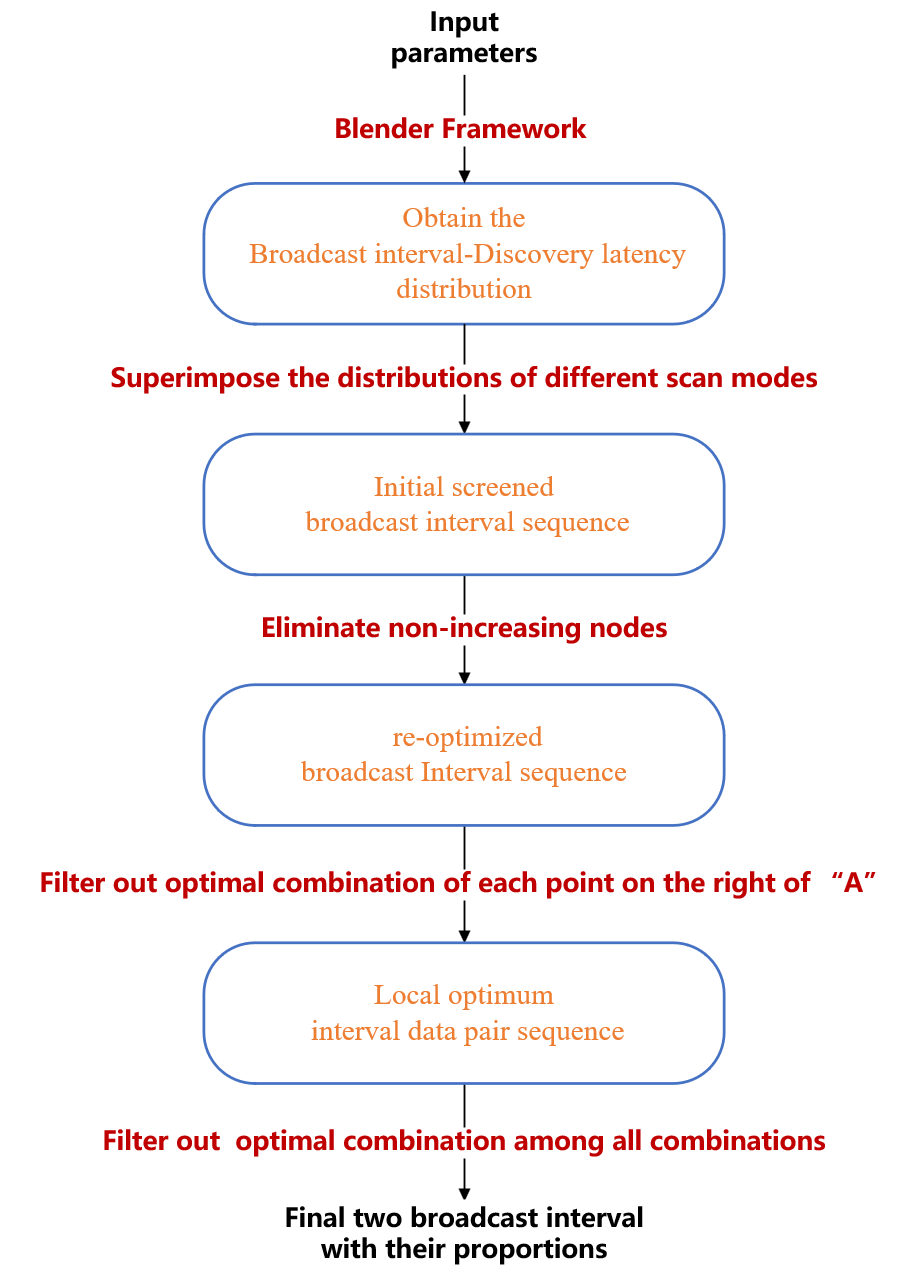}
    \caption{Working steps of the CPBIS-mechanism}
    \label{fig:enter-label}
\end{figure}

\section{Detailed Design}
Based on the previous section, this section provides detailed two-broadcast interval screening principles of the CPBIS-mechanism. By observing the distribution of discovery latency of different scan modes, we find that an arbitrary selection of one broadcast interval may result in low discovery latencies for some scan modes and high discovery latency for the others, such as b1 in Fig.11.
\begin{figure}[h]
    \centering
    \includegraphics[width=0.6\linewidth]{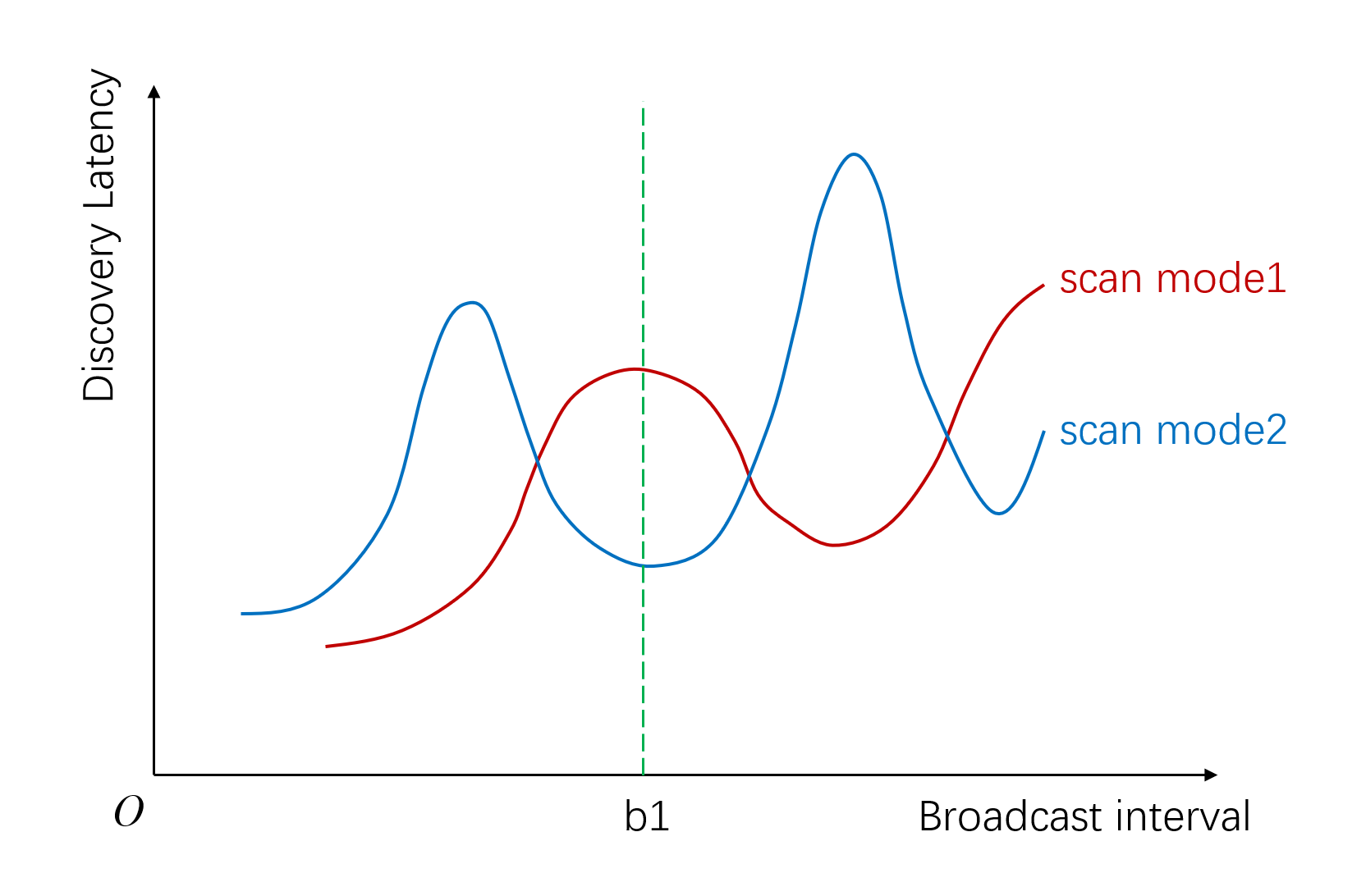}
    \caption{b1 is a local optimum interval for scan mode 2, yet it is local worst for scan mode 1.}
    \label{fig:enter-label} 
\end{figure}

\subsection{Acquisition and optimization of initial screened broadcast interval sequence}
For a single scan mode, the local optimum position sequence consists of trough points. For multiple scan modes, every point of different waveforms needs consideration. If the distributions are still viewed from a decentralized way, the problem will become increasingly complex. However, superimpose the discovery latency of different scan modes, weighted by their market share, the problem becomes simplified. As the impact of broadcast intervals to different scan modes is considered in the superimposed  discovery latency distribution, each trough point will be selected as a local optimum option following the same methodology as for an individual scan mode. As shown in Fig.12, the broadcast intervals filtered in this way avoid the peak positions in the waveform and are relatively closer to the valleys for each scan mode.
\begin{figure}[h]
    \centering
    \includegraphics[width=0.9\linewidth]{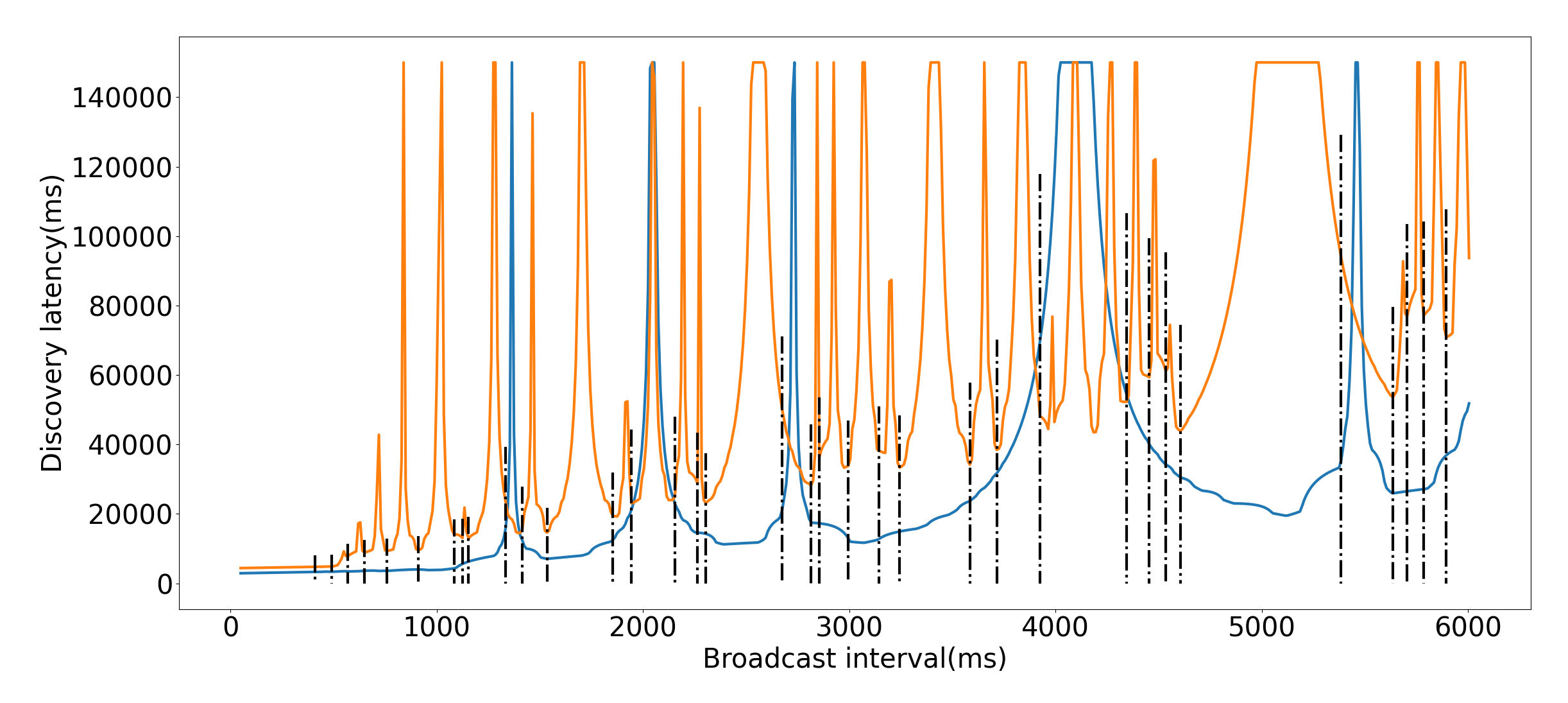}
    \caption{trough positions of superimposed distribution in original waveform}
    \label{fig:enter-label}
\end{figure}

After waveform superimposition, all trough points were selected to form the initial screened broadcast interval sequence.

After obtaining the initial screened broadcast interval sequence, further filtering is needed to remove some non-essential points. In order to reduce the computation and clarify the distribution pattern of points, we remove all non-increasing elements from right to left. As previously discussed, each non-increasing element consumes more power and exhibits higher discovery latency. Under the premise of satisfying the minimum equivalent broadcast interval constraint, their priority is lower than the first point to the right. The re-optimized broadcast interval sequence distribution appears in Fig.13 on the left. 
\begin{figure}[h]
    \centering
    \includegraphics[width=1.0\linewidth]{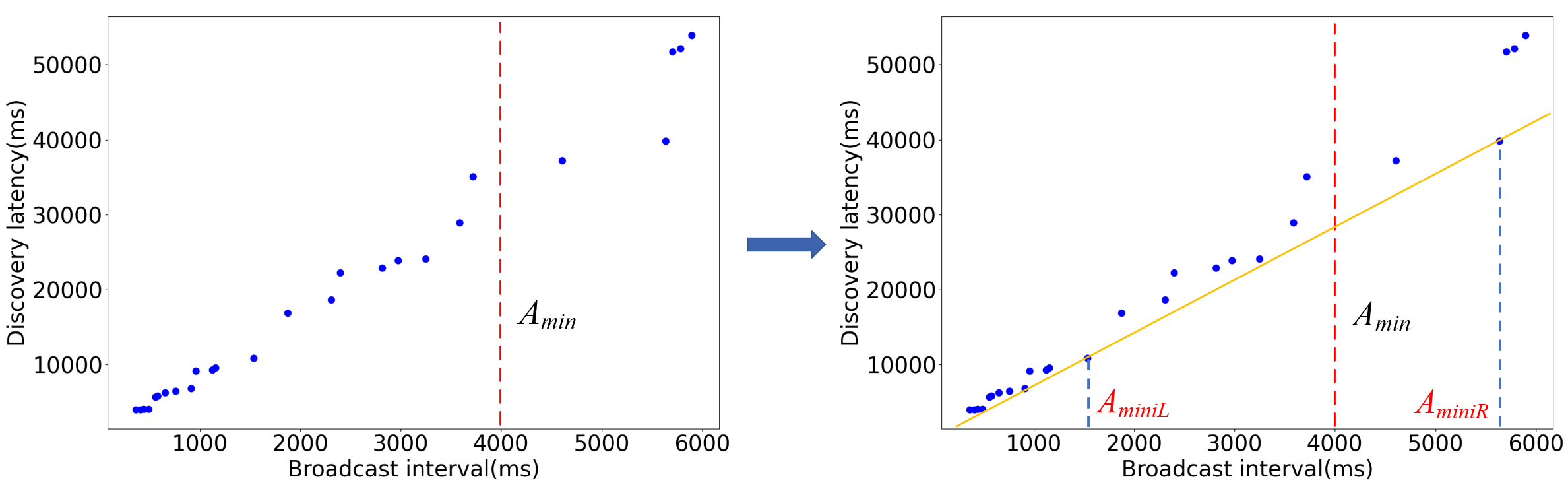}
    \caption{final two broadcast intervals}
    \label{fig:enter-label}
\end{figure}

The right picture of Fig.13 shows the final two intervals $A_{miniL}, A_{miniR}$ calculated by CPBIS-mechanism on each side of $A_{min}$, which will be clarified in the next two sections B and C.

\subsection{Analysis of re-optimized broadcast interval sequence}
Represent the point set to the left of the $A_{min}$ (e.g., $A_{min} = 4000 \text{ms}$) as $B_L$, and the point set to the right of $A_{min}$ as $B_R$. For a particular point ($A_{R1}, L_{R1}$) in the set $B_R$, there are various combinations of it with the elements in $B_L$. The following demonstrates how to select the optimal combination from all $B_L$ elements with this point. We introduce three proof procedures first:

\begin{theorem}
For any two-interval combination formed by the broadcast intervals from sets $B_L$ and $B_R$, there exists an equivalent broadcast interval $\overline{A}$ that satisfies the minimum equivalent broadcast interval constraint $\overline{A} \geq A_{min}$, and the resulting minimum discovery latency $L$ of the combination is a fixed value.
\end{theorem}
\begin{figure}[h]
    \centering
    \includegraphics[width=0.6\linewidth]{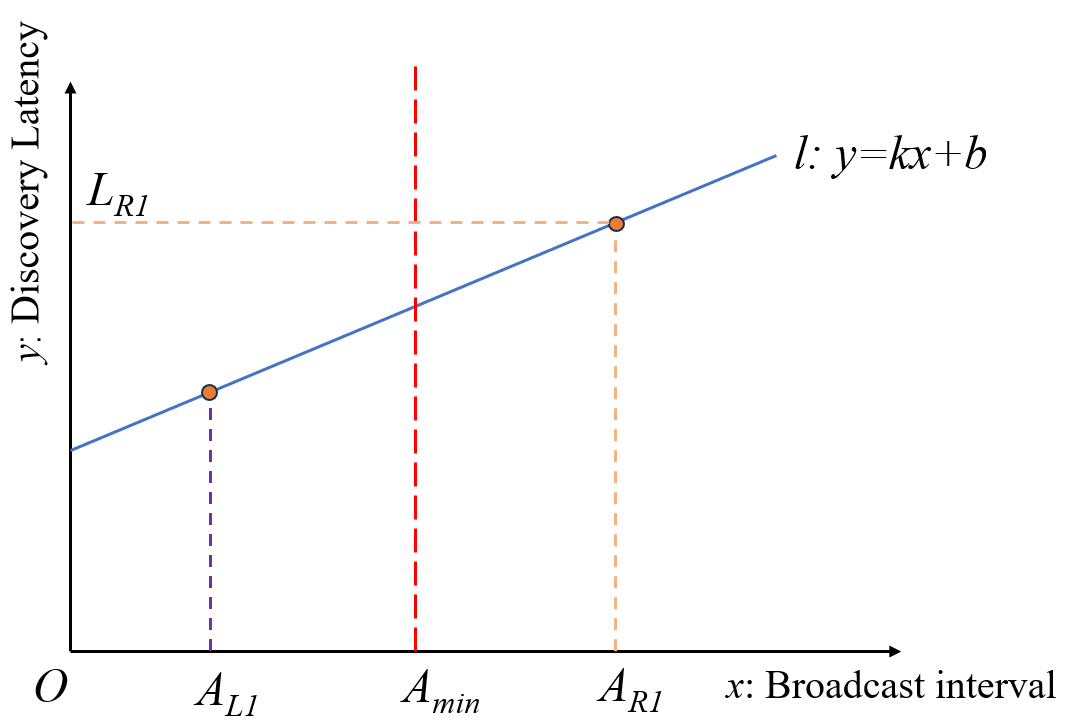}
    \caption{A random combination of interval $A_{L1}, A_{R1}$ on each side of $A_{min}$}
    \label{fig:enter-label}
\end{figure}

\begin{proof}
According to the question, shown as Fig.14, consider two intervals $A_{L1}$ in $B_L$ and $A_{R1}$ in $B_R$ on each side of $A_{min}$. The line that passes through two points is denoted as $l:y=kx+b$. To ensure the broadcast interval of this group satisfies the minimum equivalent broadcast interval constraint, we have: 
$$\alpha A_{L1} + (1-\alpha)A_{R1} \geq A_{min}$$
$$\alpha \leq \frac{A_{R1}-A_{min}}{A_{R1}-A_{L1}}$$
Therefore, when $0 \leq \alpha \leq \frac{A_{R1}-A_{min}}{A_{R1}-A_{L1}}$, the inequality holds, i.e., the constraint is satisfied.
For the combination $(A_{L1}, A_{R1})$,
\begin{equation}
\begin{aligned}
weighted\ average\ discovery\ latency\ L_1\\
= \alpha (kA_{L1} + b) + (1-\alpha)(kA_{R1} + b)\\
= k[\alpha A_{L1} + (1-\alpha)A_{R1}] + b\\
\geq kA_{min} + b
\end{aligned}
\end{equation}
When the equality in inequality(1) holds, i.e., when $\alpha$ is at its maximum, $L_1$ is minimized, which is $kA + b$.
\end{proof}

\begin{theorem}\label{th}
The minimum weighted average discovery latency generated by any two broadcast interval combination on the same straight line $l:y=kx+b$ is a fixed value.
\end{theorem}
\begin{figure}[h]
    \centering
    \includegraphics[width=0.7\linewidth]{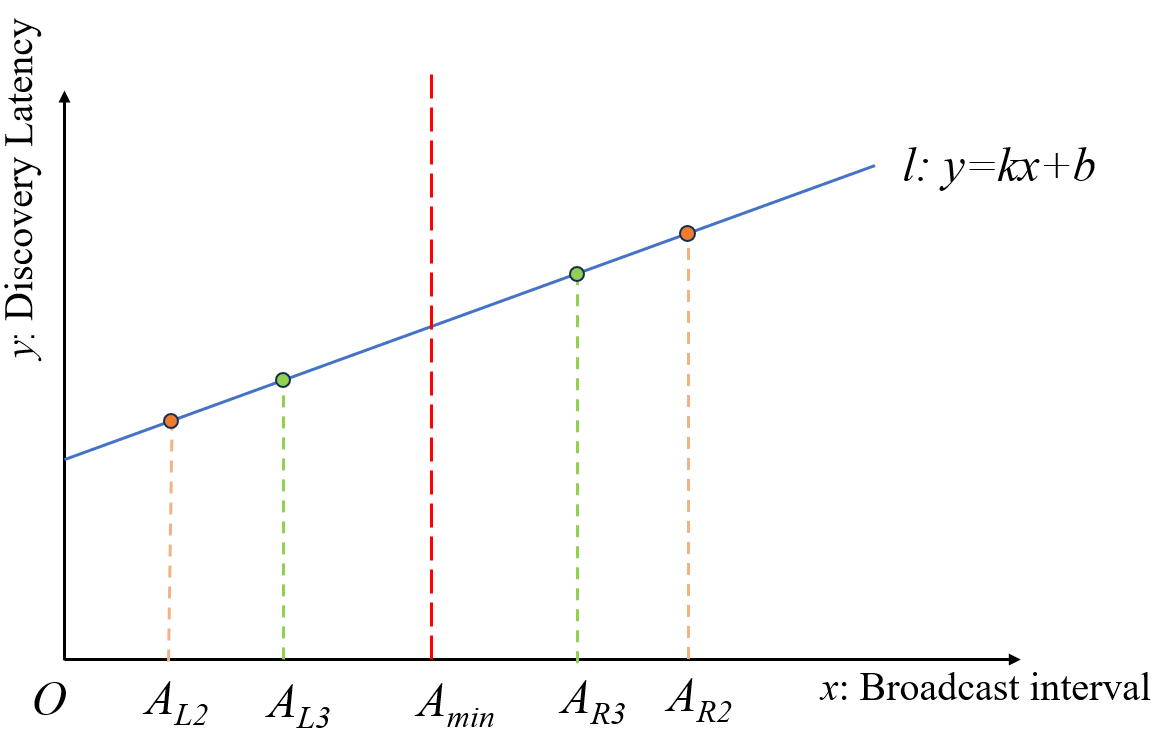}
    \captionsetup{justification=centering}
    \caption{two combinations ($A_{L2}, A_{R2}$), ($A_{L3}, A_{R3}$) on the same straight line $l$}
\end{figure}

\begin{proof}
Shown as Fig.15, suppose there are two combinations on the same straight line that satisfy the minimum equivalent broadcast interval constraint $\overline{A} \geq A_{min}$.

\begin{table}[h]
\begin{tabular}{|l|ll|ll|}
\hline
Broadcast interval combination & \multicolumn{2}{l|}{($A_{L2}, A_{R2}$)} & \multicolumn{2}{l|}{($A_{L3}, A_{R3}$)} \\ \hline
Proportion 1                   & \multicolumn{1}{l|}{$\beta$}   & 1-$\beta$  & \multicolumn{1}{l|}{$\gamma$}   & 1-$\gamma$  \\ \hline
\end{tabular}
\end{table}

According to the minimum equivalent broadcast interval constraint:
$$\begin{cases}
\beta A_{L2}+(1-\beta)A_{R2} \geq A_{min}\\
\gamma A_{L3}+(1-\gamma)A_{R3} \geq A_{min}\\
\end{cases}$$

For the combination ($A_{L2}, A_{R2}$), ($\beta, 1-\beta$),
\begin{equation}
\begin{aligned}
discovery\ latency\ L_2\\
= \beta(kA_{L2} + b) + (1-\beta)·(kA_{R2} + b)\\
= k[\beta A_{L2} + (1-\beta)·A_{R2}] + b\\
\geq kA_{min} + b
\end{aligned}
\end{equation}

In the same way,combination ($A_{L3}, A_{R3}$), ($\gamma, 1-\gamma$),
\begin{equation}
\begin{aligned}
discovery\ latency\ L_3\\
= \gamma(kA_{L3} + b) + (1-\gamma)·(kA_{R3} + b)\\
= k[\gamma A_{L3} + (1-\gamma)·A_{R3}] + b\\
\geq kA_{min} + b
\end{aligned}
\end{equation}
The minimum values of equations (4) and (5) are equal, confirming the validity of the original proposition.
\end{proof}

\begin{theorem}\label{th}
For the set $B_R$, consider any point $(A_R, L_R)$. Among all the combinations of this point with points in the set $B_L$, the combination with the largest slope corresponds to the minimum weighted average discovery latency.
\end{theorem}
\begin{figure}[h]
    \centering
    \includegraphics[width=0.7\linewidth]{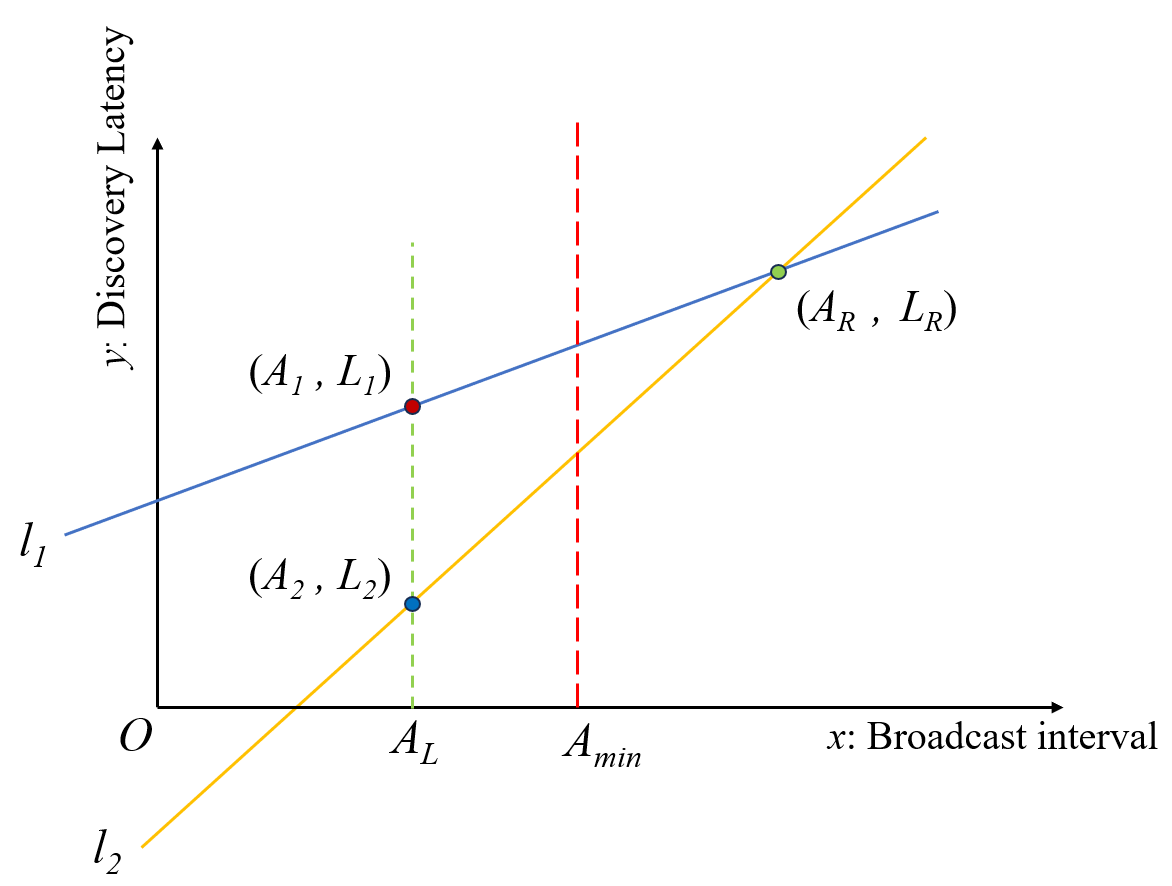}
    \captionsetup{justification=centering}
    \caption{Two straight lines passing through ($A_R, L_R$)}
    \label{fig:enter-label}
\end{figure}
\begin{proof}
Shown as Fig.16, draw any two lines (with slopes $k > 0$) passing through the point ($A_R, L_R$). Suppose there exists a point $A_L$ to the left of $A_{min}$. The discovery latencies at the intersections of the vertical line through $A_L$ with these two lines are $L_1$ and $L_2$ from top to bottom.

1). Slope comparison:
$$l_1:k1=\frac{L_R-L_1}{A_R-A_L}$$
\[l_2:k2=\frac{L_R-L_2}{A_R-A_L} \]

Since $L_1 \textgreater L_2$, thus $k1 \textless  k2$.

2). comparison of sum of discovery latency:

Suppose that for the combination ($A_L, A_R$), there exists a proportion $(c, 1 - c)$ such that the constraint $\overline{A} \geq A$ is satisfied, then:

$$ for\ l_1:L_{1total}= c L_1+(1-c)L_R $$
\[for\ l_2:L_{2total}=c L_2+(1-c)L_R\]
$$L_{1total}-L_{2total}=c (L_1-L_2) \textgreater 0 $$

Furthermore, with $k_1 < k_2$ and in conjunction with Theorem 2, it follows that even if the two points on the left side of $A_{min}$ forming a combination with $A_R$ are not on the same vertical axis, the combination with the larger slope still results in a smaller discovery latency.
\end{proof}

Using a similar method as in Theorem 3, the following conclusion can be obtained: For the set $B_L$, consider any point $(A_L, L_L)$. Among all the combinations of this point with points in the set $B_R$, the combination with the smallest slope corresponds to the minimum weighted average discovery latency. We call this conclusion 4.

Based on Theorems 2 and 3, we process the re-optimized interval sequence as shown in Figure 16:

Use the minimum broadcast-interval $A_{min}$ as the dividing line, start from the first point ($B_{R1}, L_{R1}$)on the right side of $A_{min}$, compare the slope of this point with all points on the left side of $A_{min}$ (Fig.17(a)), use the theorem 3, the combination with the largest slope is the optimal combination formed with this point, among all the straight lines passing through the point ($B_{R1}, L_{R1}$), a combination obtains the minimum weighted average discovey latency can be found, shown as Fig.17(b).
\begin{figure}[h]
    \centering
    \subfloat[]{
    \centering
    \label{}
    \includegraphics[width=0.24\textwidth]{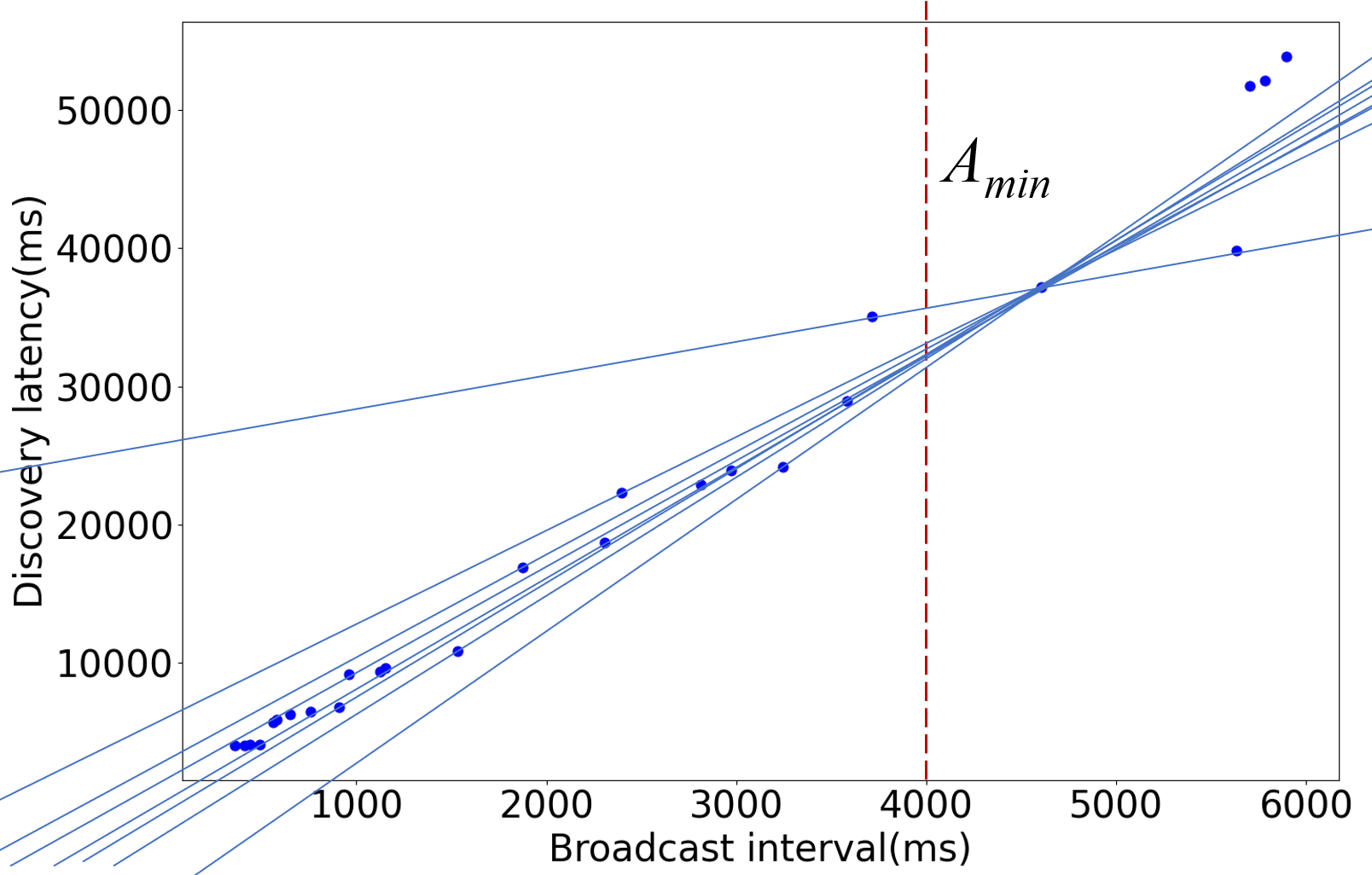}}
    \subfloat[]{
    \label{}
    \includegraphics[width=0.24\textwidth]{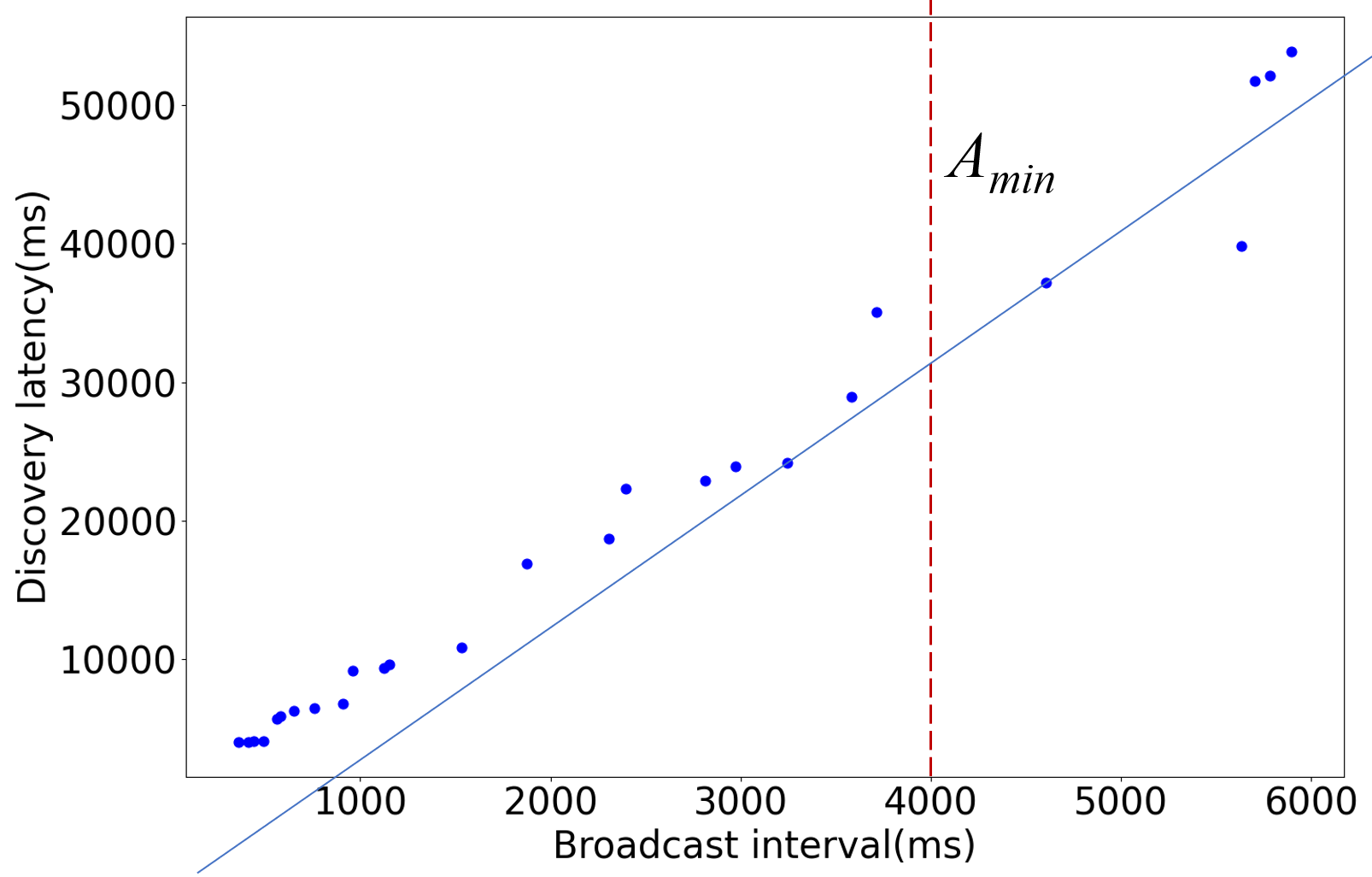}}
    \caption{(a): Combination of straight lines passing through the point ($B_{R1},L_{R1}$) - corresponding to various broadcast interval combinations. (b): Optimal combination with broadcast interval 4605 (3245,4605)}
\end{figure}

\subsection{Optimal interval data pair of all combinations}
Part B discusses the case of one point in set $B_R$, while Part C expands the study to the entire $B_R$, searching for the optimal result with the minimum weighted average discovery latency among all combinations formed by points in $B_L$ and $B_R$
\begin{figure}[h]
    \centering
    \subfloat[]{
    \centering
    \label{}
    \includegraphics[width=0.25\textwidth]{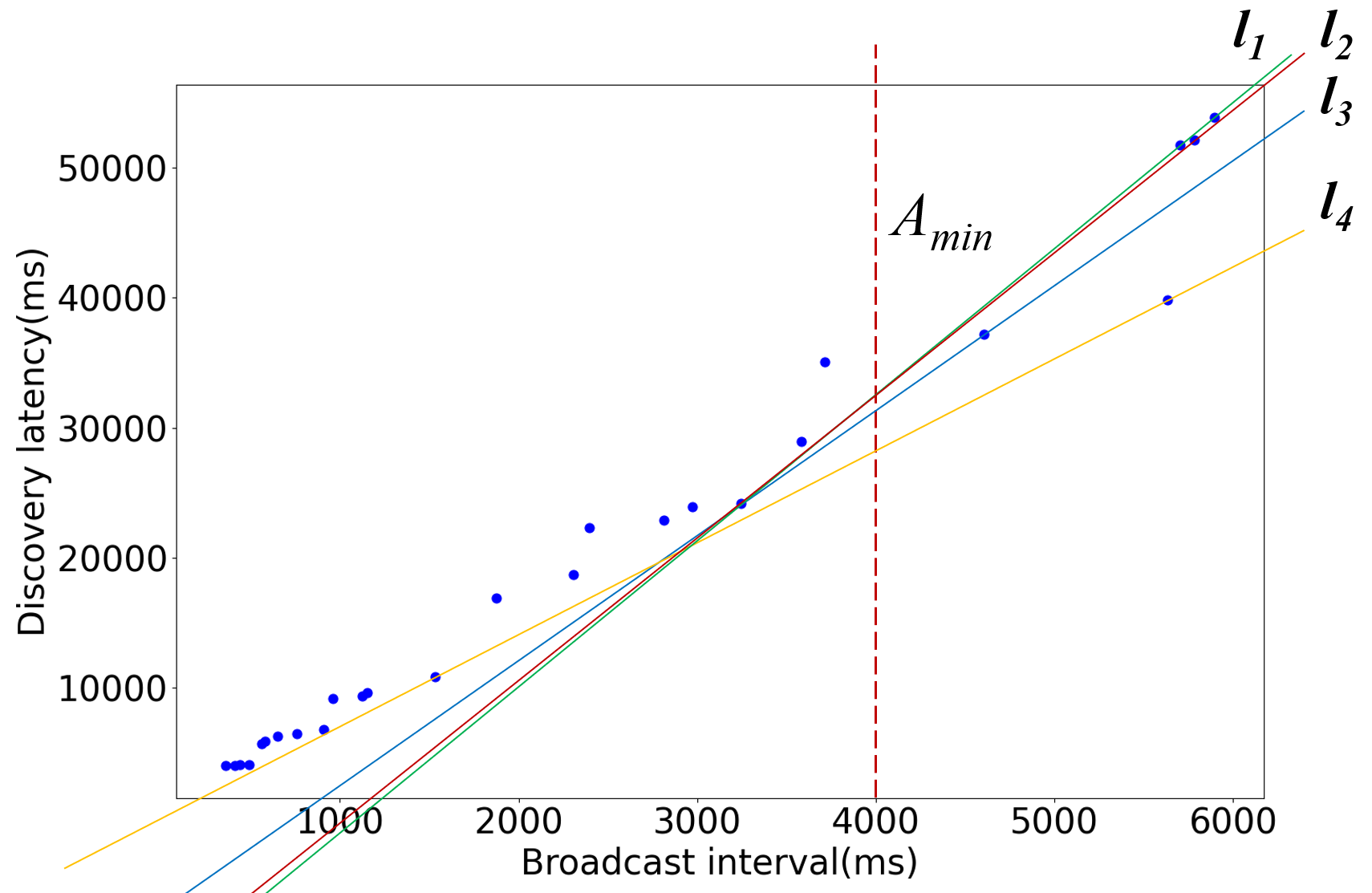}}
    \subfloat[]{
    \label{}
    \includegraphics[width=0.25\textwidth]{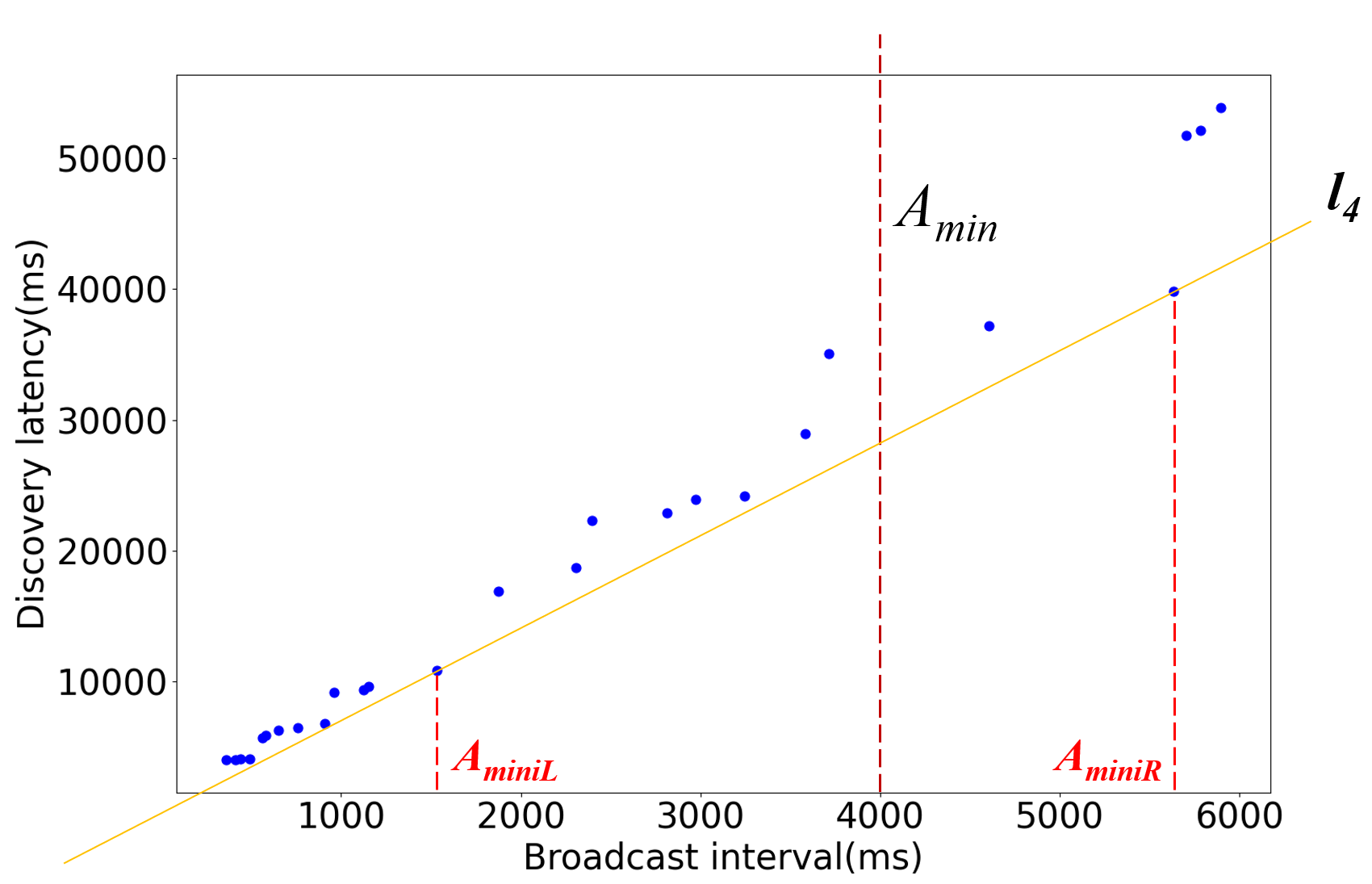}}
    \caption{(a): Local optimum interval data pairs. (b): Optimal combination with broadcast interval $A_{miniL}, A_{miniR}$}
\end{figure}

Continue with the work of Part B, when the last point in $B_R$ is traversed, the optimal broadcast interval combination belonging to each point in $B_R$ has been derived. These interval data pairs form a local optimum interval data pair sequence. We name the straight lines formed by each combination as $l_1, l_2...l_n$ in the order of their intersections with the  dividing line from top to bottom as in the Fig.18(a). The next is to compare these combinations and select the most appropriate combination from the local optimum interval data pair sequence.

First, any line from $l_1$ to $l_n$ intersects with all the remaining lines, it is impossible for parallel lines to exist. Assuming parallel lines exist, one must be above and the other below. In this case, the line above have to select the lower point on the left side of $A_{min}$  to form a larger slope, resulting in an intersection.

Second, the intersection must be to the left of $A_{min}$. Assume that the intersection of the two lines is to the right of $A_{min}$, then the line with the larger slope must pass through a point in $B_L$, and the other line passing through this point would have a slope greater than the original one. which means the original slope is not the largest slope. Therefore, any two lines have an intersection point that lies to the left of $A_{min}$. Combine with conclusion 4, simply select the data pair with the smallest slope. The yellow line $l_4$ is the desired line .

The two broadcast interval is shown as Fig.18(b) $A_{miniL}, A_{miniR}$. The proportion of the two broadcast intervals can be derived by substituting $A_{miniL}, A_{miniR}$ into $ \overline A = \delta A_1 + (1 -\delta) A_2 $ with $\overline A = A_{min}$.

the algorithm of CPBIS-mechanism is like this:
\begin{algorithm}[h]
\caption{Find Optimal Broadcast-Interval Data Pairs}
\SetKwInOut{Input}{Input}
\SetKwInOut{Output}{Output}
\Input{Different Broadcast Interval-Discovery Latency Distributions}
\Output{The optimal broadcast-interval data pairs}

Superimpose the discovery latency distributions\;
$B1[\,] \leftarrow$ Select out the trough points of the superimposed distribution\;
\For{$i \leftarrow \text{len}(B1)-1$ \KwTo $0$}{
    Delete the non-increasing points from $B1[\,]$\;
}
\For{$j$ \textbf{in} $B1[\,]$}{
    \If{$j < A$}{
        $B_L[\,] \leftarrow j$\;
    }
    \Else{
        $B_R[\,] \leftarrow j$\;
    }
}
\For{$m$ \textbf{in} $B_R[\,]$}{
    \For{$n$ \textbf{in} $B_L[\,]$}{
        Compare the slope computed with $m$, $n$\;
        $a \leftarrow$ The pair with the largest slope\;
    }
    $B2[\,] \leftarrow a$\;
}
\For{$k$ \textbf{in} $B2[\,]$}{
    Compare the slope computed with $k$\;
    $A_{miniL}, A_{miniR}\leftarrow$ The pair with the smallest slope\;
}
\end{algorithm}

\section{Performance evaluation}

In order to experimentally verify our CPBIS-mechanism, we made the following deployments:

The experiment is a BLE neighbor discovery test based on the Nordic nRF52832 chips, prototyped as a joint working model of Apple's AirTag with "Find My" Network to find lost items. We chose Keil uvision5\cite{KeilMDK5} as the hardware code editing platform. Adjustments to both the broadcast interval and scan interval of the nRF52832 chips are realized using this platform. We use the nRF5 SDK 17.0.2\cite{nRF5SDK} officially provided by Nordic and modify it to realize the required functions. The protocol stack is burned into the chip using nRFgo studio\cite{nRFgoStudio} provided by Nordic. We use J-Link V7.96\cite{JLinkDriver} and a J-link emulator(Fig.19(b)) to program and debug Keil projects.As the J-link RTT Viewer can not save scan logs in real time, it periodically clears data. We have identified the Real Time Transfer Tool (RTT-T)\cite{RTT-T}, provided by Liuhao, to preserve chip logs.

The BLE neighbor discovery experiment requires a "Scanner" and a "Advertiser". We chose two nRF52832 development boards(Fig.19(a)) as the Advertiser and the Scanner respectively. The Advertiser simulates the lost BLE tag, and the Scanner simulates the finder device. The nRF52832 chip is chosen because the Bluetooth module used in the real prototype Air Tag is nrf52832. By adjusting the broadcast mode of the chip, we simulate the situation that the offline tag broadcasts its signal to the surroundings when it is lost. Based on previous research\cite{li2023design}, the broadcast modes we set include a single broadcast interval, alternating broadcasts with the same proportion of two broadcast intervals, and two broadcast intervals with different proportions. As for the Scanner, we adjust its scan mode to simulate the various BLE scan devices that join the "Find My" network. Additionally, when setting the scan mode on the Keil uvision5, we add a scan report event to it. The event includes the names of the broadcast devices received, enabling us to search the logs and determine if the Advertiser was successfully found.
The experimental deployment is shown in Fig.20.
\begin{figure}[h]
    \centering
    \subfloat[]{
    \centering
    \label{}
    \includegraphics[width=0.24\textwidth]{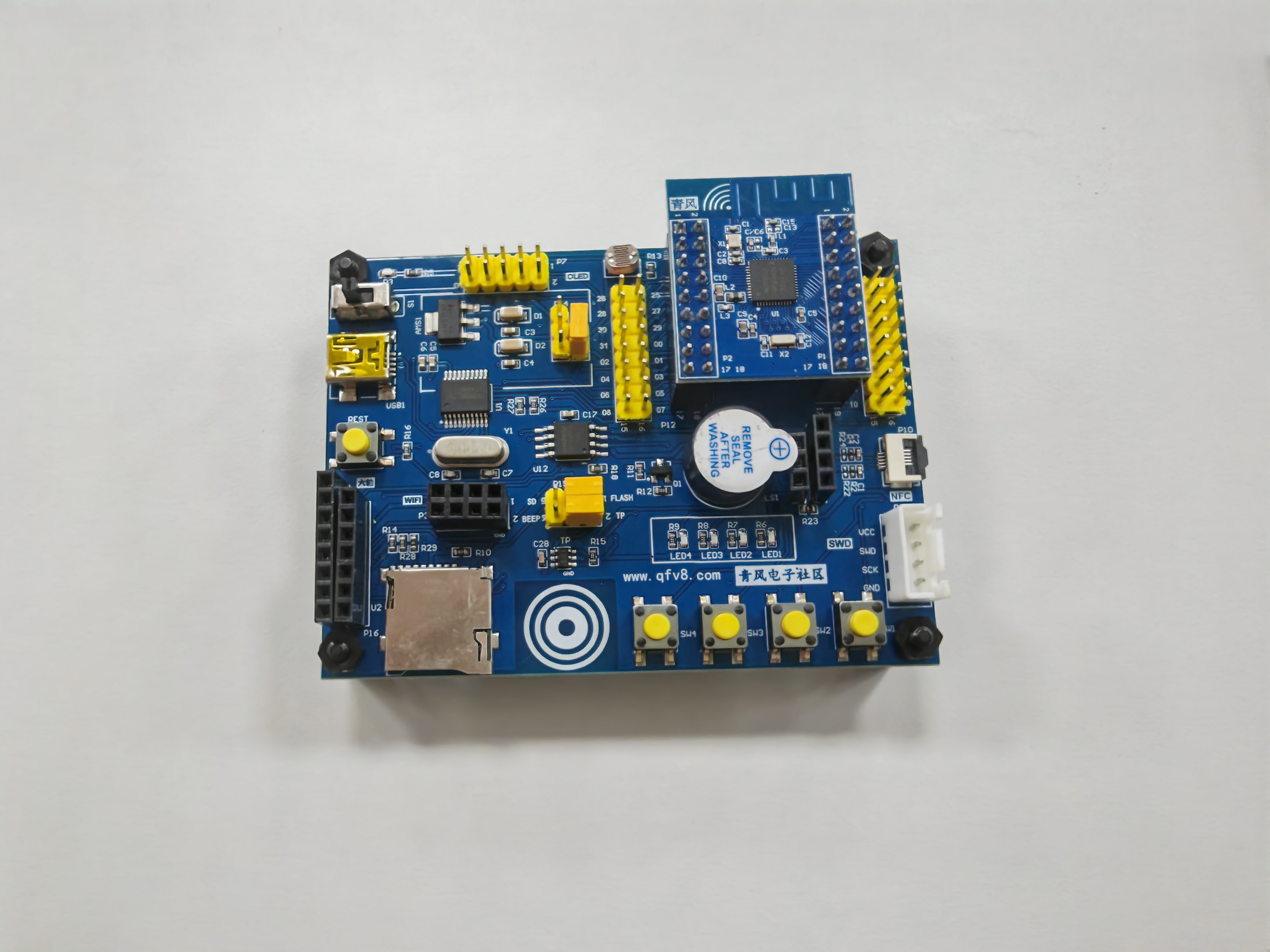}}
    \subfloat[]{
    \label{}
    \includegraphics[height=0.24\textwidth, angle=90]{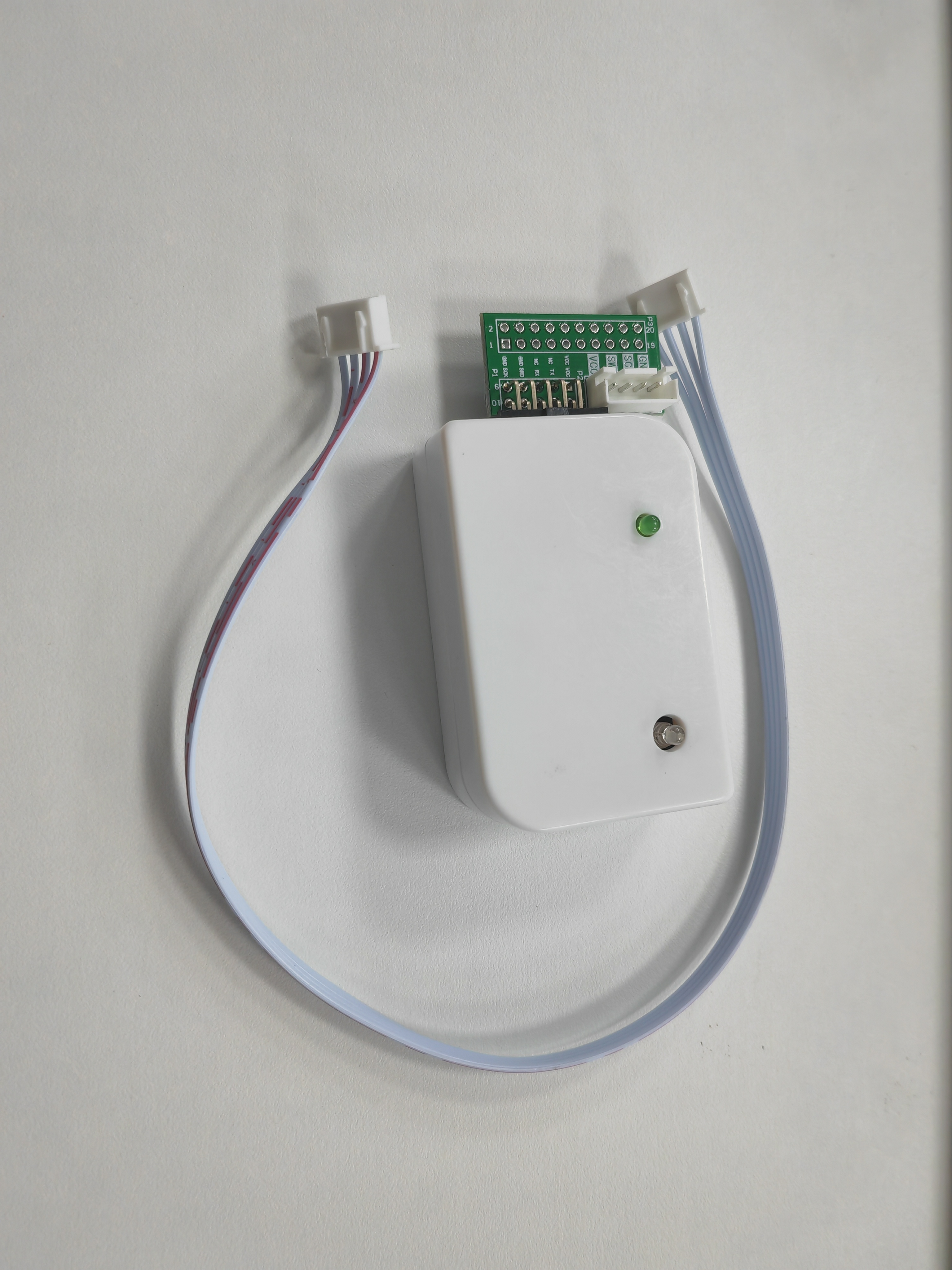}}
    \caption{(a): nRF52832 development board, there are two boards in total, one is used as the "Scanner" and the other is used as the "Advertiser". (b): J-Link emulator}
\end{figure}

\begin{figure}[h]
    \centering
    \includegraphics[width=0.6\linewidth]{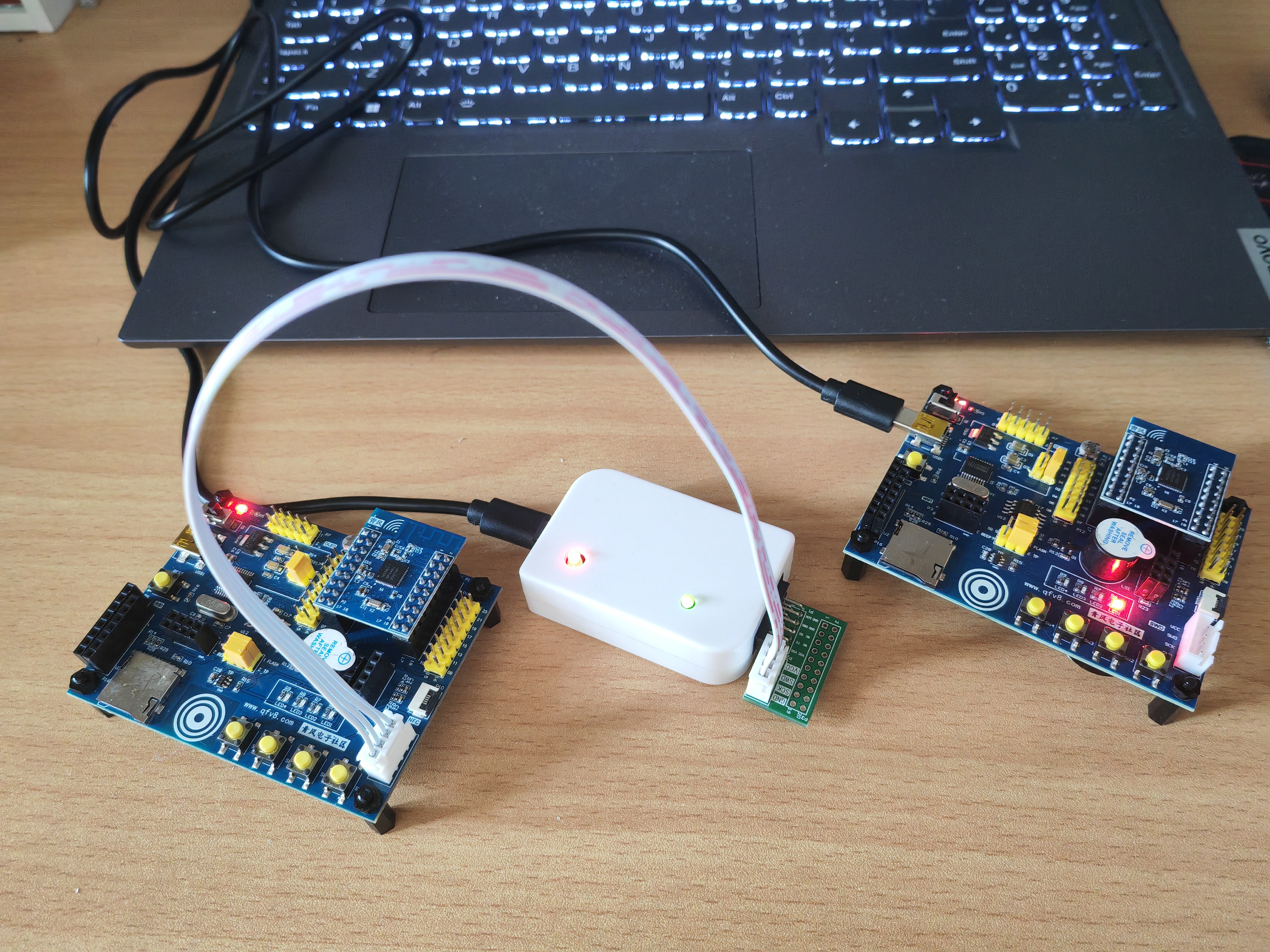}
    \caption{CPBIS testing in BLE neighbor discovery, with Scanner on the left and Advertiser on the right}
    \label{fig:enter-label}
\end{figure}
The experimental objective is to explore the success rate of discovering the tag within a limited time and space using CPBIS, and compare it with previous results. And record the discovery latency and compare the weighted average discovery latency of different experimental groups to validate our conclusion.

To compare with previous work, we maintained the same premise as in \cite{li2023design}. Suppose there are two scan modes on the market. The first scan mode has a scan interval of 4096 ms and a scan window of 1024 ms; the second scan mode has a scan interval of 5120 ms and a scan window of 512 ms. Both scan modes have an market share of 50\%.

We set three broadcast modes: A) a single broadcast interval of 4600 ms, B) 2980 ms and 5620 ms alternately broadcast for 20 s, and C) 1535 ms and 5645 ms alternately broadcast for 16 s and 24 s, respectively, obtained by the CPBIS-mechanism. The duration of each interval in modes B and C represents its proportion in the two-interval broadcast mode. Settings for the broadcast mode is presented in Table I.

\begin{table}[h]
	\caption{parameter settings}
        \renewcommand\arraystretch{1.2}
	\centering
        \resizebox{\columnwidth}{!}{
	\begin{tabular}{|c|c|c|}
		\hline
		\diagbox{broadcast mode}{broadcast interval}&the first interval&the second interval\\
		\hline
		A. a single broadcast&4600ms for 40s&NONE\\
		\hline
		B. alternative broadcast&2980ms for 20s&5620ms for 20s\\
		\hline
		C. CPBIS broadcast&1535ms for 16s&5645ms for 24s\\
		\hline
	\end{tabular}
        }
\end{table}
The experimental procedure is as follows:

In reality, since the situations of people passing by the offline tag are random, a scan device may enter the tag's broadcast range at any stage of the broadcast. Simulating pedestrians repeatedly entering the broadcast range of the tag in experiments would be time-consuming, and the broadcast range of the tag is not a precise value, which could introduce significant error. Therefore, within the range ensuring that the broadcast signal of Advertiser can be received, we randomly activate the Scanner to simulate pedestrians randomly entering the tag's broadcast range.

We limit the duration of one BLE neighbor discovery process to 40 seconds and group the experiments based on the two scan modes. For the three broadcast modes A,B and C. We take 10 BLE neighbor discovery processes as a small group and keep increasing the number of small groups until the discovery success rate stabilize.

The first group is T/W = 4096 ms / 1024 ms, i.e., 4096 ms as the scan interval and 1024 ms as the scan window. First, the broadcast mode A was used for the Advertiser. When the broadcast is turned on, 30 neighbor discovery experiments are conducted randomly. Connecting the Scanner to the J-link emulator and using the J-link RTT-T tool to record and save the scan report events. In these 30 experiments, the Scanner discovered the presence of the Advertiser with an average discovery latency of 13.273s. Then, mode B was used as the Advertiser's broadcast mode. Again, 30 neighbor discovery experiments were conducted, and the Scanner successfully found the Advertiser within the time limit each time. This time, the scan report event concluded that the average discovery latency in broadcast mode B was 7.091s. Finally, for the broadcast mode C. The presence of the Advertiser was discovered in all 30 neighbor discovery experiments with an average discovery latency of 6.627s, the lowest among the three broadcast modes.

The second group, T/W = 5120 ms / 512 ms, Ultimately, each group conducted a total of 60 neighbor discovery experiments. For broadcast mode A with a single broadcast interval of 4600 ms, the success rate of 60 neighbor discoveries is 81.6\%, and the average discovery latency is 17.322 s. For broadcast mode B with 2980 ms and 5620 ms alternately broadcasting for 20s, the success rate of discovery is 71.6\%, and the average discovery latency is 14.045s. For broadcast mode C with 1535 ms and 5645 ms broadcasting for 16s and 24s respectively, the success rate of discovery is 98.3\%, and the average discovery latency is 14.562s. 

Unlike equation(2) in Section III.B, the weighted average discovery latency in the actual experiments is computed by adding up the results in two scan modes by weight (market share), the weighted average discovery success rate is calculated in the same way. Therefore, the weighted average discovery success rates of the three broadcast modes A, B, and C are 90.8\%,85.8\% and 99.15\%, respectively,shown as Fig.21. The weighted average discovery latencies of the three broadcast modes A, B, and C are calculated to be 15.298s,10.568s and 10.595s, respectively. The experimental data on discovery latency is presented in Table II.

\begin{table}[h]
	\caption{Discovery latency for different broadcast modes}
        \renewcommand\arraystretch{1.2}
	\centering
        \resizebox{\columnwidth}{!}{
	\begin{tabular}{|c|c|c|c|}
		\hline
		\diagbox{broadcast mode}{discovery latency}{scan modes:T/W}&4096ms/1024ms&5120ms/512ms&weighted average latency\\
		\hline
		A. a single broadcast&13.273s&17.322s&15.298s\\
		\hline
		B. alternative broadcast&7.091s&14.045s&10.568s\\
		\hline
		C. CPBIS broadcast&6.627s&14.562s&10.595s\\
		\hline
	\end{tabular}
        }
\end{table}

\begin{figure}[h]
    \centering
    \includegraphics[width=0.7\linewidth]{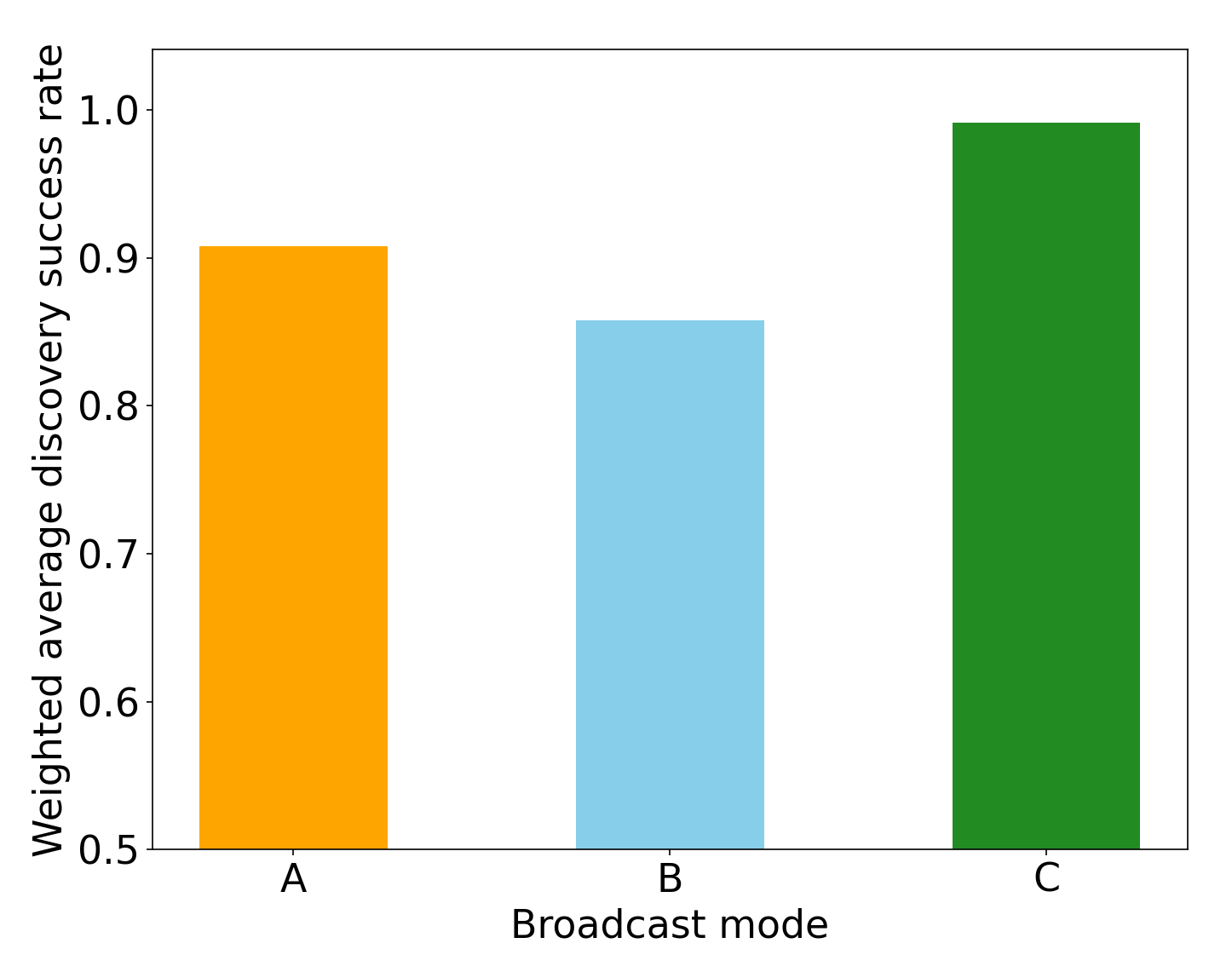}
    \caption{weighted average discovery success rate for different broadcast modes}
    \label{fig:enter-label}
\end{figure}
\section{Related Work}
The techniques applied in BLE offline finding network are mainly focused on BLE neighbor discovery. Modeling analysis of the process was available early on. An analytical model is proposed in paper\cite{liu2012modeling} for 3-channel-based neighbor discovery in BLE networks. It inspired later BLE NDP studies. Paper\cite{luo2020neighbor} proposed a theoretical model based on Chinese Remainder Theorem(CRT), and analyzed the discovery  latency performance of BLE NDP using this model. Paper\cite{kindt2019optimal} found the worst discovery latency boundary of the “broadcast interval-discovery latency” distribution. In\cite{ding2022blender}, a simulator “Blender” is proposed to generate a CDF graph of the discovery latency for a given parameter configuration. Further extension can obtain the “broadcast interval-discovery delay” distribution for a fixed scan mode, which is the basis of CPBIS.

Offline finding network is currently in use. Huawei, Apple, and Samsung all have related businesses, which shows its great promise in finding lost items. The entire workflow is relatively complete, but the detailed setting of parameters in the workflow is still under study. How to optimize the discovery process by setting some key parameters like broadcast mode, and then improve the user experience is still a problem to be solved. While CPBIS gives a complete solution of broadcast interval screening.

The current configuration of BLE tag in offline finding network is relatively simple, e.g., the broadcast interval of Apple's Air Tag in lost state is fixed at 2000ms\cite{AirTagReverse}. This can lead to unsatisfactory results in BLE neighbor discovery in practical applications, resulting in excessively high discovery latency and low discovery success rate.

In order to achieve an ideal NDP performance, parameter configurations such as broadcast interval, scan interval, and scan window are particularly important. In the paper\cite{luo2019ble}, after modeling the NDP process, some key parameters in the BLE NDP process are given recommended configurations, in addition to the above three common parameter configurations, the range of values for the broadcast interval, and the random broadcast delay (adv delay) are also restricted. The article mainly considers the impact of broadcast interval on NDP discovery latency, and finds that there are a large number of peaks in the “broadcast interval-discovery latency” distribution, and it is necessary to be careful to avoid parameters that lead to peaks in the actual parameter configuration. This is also the optimization goal of CPBIS.

The BLE broadcast mode is more unified and controllable in offline finding network. That’s why researchers often reduce the discovery latency of the BLE NDP by optimizing the tag broadcast mode. In order to optimize the discovery latency in BLE NDP, a method to determine the optimal broadcast interval for BLE broadcast devices is proposed in the paper\cite{shan2018advertisement}. to minimize the time for a scan device to discover the surrounding broadcast devices. However, this study focuses on the case that there are many broadcast devices and all of them are to be discovered, whereas the BLE offline finding network is the case that there is one broadcast device and an unkonown number of scan devices.

In order to find out the suitable broadcast mode for multiple scan modes, a broadcast mode “ElastiCast” is conceived in\cite{li2023design}, which consists of two types of broadcast patterns, namely, Single Broadcast Pattern (ElastiCast-SBP) and Alternation Broadcast Pattern. (ElastiCast-ABP). ElastiCast has given a clear screening scheme for single broadcast interval selection. While more than two broadcast intervals are still under study. CPBIS is a complete solution for screening two broadcast intervals and determining their proportions.

\section{Conclusion}
This paper proposes a two-broadcast-interval screening scheme for BLE offline finding network. With CPBIS-mechanism, the broadcast mode elected under multiple Bluetooth scan modes (T/W) achieves relatively low discovery latency for all scan modes, which improves the success rate of BLE neighbor discovery within a limited period of time. In order to validate our CPBIS mechanism, we tested it on the nRF52832 chip, and compared it with the previous work. The weighted average discovery latency within 40s is 10.595s, the weighted discovery success rate is the highest among the three broadcast modes under the same test condition. This confirms the effectiveness of CPBIS-mechanism.

\end{document}